\RequirePackage{fix-cm}

\documentclass{svjour3}                     
\smartqed  
\usepackage[table,xcdraw]{xcolor}
\usepackage[font=small,labelfont=bf,tableposition=top]{caption}
\usepackage[ruled,linesnumbered,vlined]{algorithm2e}
\usepackage{amsmath}
\usepackage{amsfonts}
\usepackage{amssymb}
\usepackage{graphicx}
\usepackage{mathtools}
\usepackage{ntheorem}
\usepackage{tikz}
\usepackage{multirow}
\usepackage{rotating}
\usepackage{xcolor}
\usepackage{thmtools}
\usepackage{url}
\usepackage[breaklinks]{hyperref}
\usepackage{breakurl}
\usepackage{times}
\usepackage[normalem]{ulem}
\usepackage{esvect}

\useunder{\uline}{\ul}{}

\DeclareCaptionLabelFormat{andtable}{#1~#2  \&  \tablename~\thetable}

\newcommand{\acro}[1]{{\sc \lowercase{#1}}}

\newcommand{\threepartdef}[6]
{
	\left\{
		\begin{array}{lll}
			#1 & \mbox{if } #2 \\
			#3 & \mbox{if } #4 \\
			#5 & \mbox{if } #6
		\end{array}
	\right.
}

\definecolor{javi}{RGB}{0,0,255}

\newtheorem{mydef}{Definition}

\renewcommand\thmcontinues[1]{Continued}

\DeclareCaptionLabelFormat{andtable}{#1~#2  \&  \tablename~\thetable}

\journalname{Auton Agent Multi-Agent Syst}

\begin{document}

\title{Synthesising Evolutionarily Stable Normative Systems
}

\titlerunning{Synthesising Evolutionarily Stable Normative Systems}        

\author{Javier Morales \and Michael Wooldridge \and \\ Juan A. Rodr\'iguez-Aguilar \and Maite L\'opez-S\'anchez}


\institute{
Javier Morales and Michael Wooldridge
\at Department of Computer Science, University of Oxford. Oxford, United Kingdom\\
\email{javier.morales@cs.ox.ac.uk}\\
\email{michael.wooldridge@cs.ox.ac.uk}
\and
Juan A. Rodr\'iguez-Aguilar \at Artificial Intelligence Research Institute (IIIA-CSIC). Campus de la UAB, Bellaterra, Spain\\
\email{jar@iiia.csic.es}
\and
Maite L\'opez-S\'anchez \at 
Department of Applied Mathematics and Computer Science, Universitat de Barcelona. Barcelona, Spain\\
\email{maite@maia.ub.es}
}

\date{Received: date / Accepted: date}

\maketitle

\begin{abstract}
Within the area of multi-agent systems, normative systems are a widely used framework for the coordination of interdependent activities. A crucial problem associated with normative systems is that of synthesising norms that effectively accomplish a coordination task and whose compliance forms a rational choice for the agents within the system. In this work, we introduce a framework for the synthesis of normative systems that effectively coordinate a multi-agent system and whose norms are likely to be adopted by rational agents. Our approach roots in evolutionary game theory. Our framework considers multi-agent systems in which evolutionary forces lead successful norms to prosper and spread within the agent population, while unsuccessful norms are discarded. The outputs of this evolutionary norm synthesis process are normative systems whose compliance forms a rational choice for the agents. We empirically show the effectiveness of our approach through empirical evaluation in a simulated traffic domain.
\end{abstract}


\section{Introduction}
\label{sec:intro}


Within both human and agent societies, normative systems (norms) have been widely studied as mechanisms for coordinating the interplay between autonomous entities \cite{boella2006introduction,MASfoundations09}. Given a society, norms can resolve coordination problems by guiding the decision-making of its individuals, restricting their behaviour once some preconditions are fulfilled. 

In the literature on norm research, normative systems are typically represented as sets of \textit{soft} constraints on the behaviour of agents, who can autonomously decide whether or not to comply with them. Often, agents face a choice between \textit{norm compliance}, which allows them to achieve the social welfare at an individual cost, and infringement, which enables them to achieve better individual results at the cost of jeopardising social welfare. Thus, norm compliance is a concern in the area of normative systems, which could be summarised by means of the following questions:
\begin{enumerate}
\item Will the individuals of a society comply with the norms of a given normative system?
\item If not, what type of normative systems will they comply with? How to synthesise them? 
\end{enumerate}

Accordingly, much work in the literature in norm research has focused on the problem of norm compliance \cite{sethi2005compliance,governatori2010norm,bicchieri2009compliance}, and particularly on how to synthesise norms that discourage non-compliant behaviour \cite{bicchieri2009compliance,bicchieri2010behaving}. Along these lines, some works like \cite{axelrod1986evolutionary,sethi1996evolutionary,shoham1997emergence,azar2004sustains,sugawara2011emergence} have taken inspiration on the framework of evolutionary game theory (EGT) \cite{smith1988evolution} to understand the process whereby societies come to adopt norms. They consider a setting in which agents repeatedly play a game (e.g., the Prisoner's Dilemma \cite{rapoport1965prisoner}) by using different strategies. Strategies that are seen to be successful prosper and spread within the agent society through an evolutionary game theoretic process whereby agents tend to adopt successful strategies with higher probabilities than unsuccessful ones. A norm is regarded as a behavioural regularity that emerges within the society: a norm is said to have been established once a majority of agents adopt the same strategy, which everyone prefers to conform, on the assumption that everyone else does. Such a strategy has the property that complying with it is a rational choice for the agents, and hence it is said to be \textit{evolutionarily stable} (ESS): once the agents adopt it, no agent can benefit from deviating. 


Although EGT has been proven to be useful to predict which norms can be evolutionarily stable, most of the works on EGT and norms make strong assumptions that are inconvenient when synthesising norms for multi-agent systems (MAS). First, they consider that agents play a \textit{single} game \--- and hence, a single norm can be synthesised. In fact, agents in a MAS typically engage in a wide variety of interaction situations (games) that may require different norms. As an example, humans have designed different traffic rules to coordinate drivers in a variety of situations (e.g., when entering a junction, or when overtaking a vehicle). Hence, MAS coordination may require employing \textit{sets of norms} instead of a single one. Second, these works assume a deterministic setting in which the game that the agents can play along with its payoffs are known beforehand. However, some systems may have a certain degree of non-determinism that makes it impossible to assume the outcomes of agents' interactions. For instance, one may not be able to ensure that a car will not have an accident once it stops at a red light \--- the brakes may fail, or another car may hit it from behind. 


Against this background, this paper contributes to the state of the art by introducing a framework for the synthesis of \textit{evolutionarily stable normative systems} (ESNS) for non-deterministic settings. Our framework incorporates ideas from EGT. It carries out an evolutionary process whereby the agents of a MAS can ultimately adopt sets of coordination norms that are evolutionarily stable (and hence, whose compliance is rational for them). Our framework assumes that the potential coordination situations and their outcomes are unknown beforehand. Agents are permitted to interact, and our framework discovers these situations at runtime, modelling them as games, and empirically computing their payoffs. Norms that successfully coordinate the agents prosper and spread, and agents ultimately adopt evolutionarily stable norms. We provide empirical evaluation of our framework in a simulated traffic domain, and we show that it can synthesise ESNSs that avoid car collisions.

Our framework provides a valuable tool for decision support for policy makers. Given a society, it opens the possibility of:
\begin{enumerate}
\item Synthesising an ESNS that successfully accomplishes a coordination task.
\item Predicting whether the agents will comply with a given normative system or not, and if they do not, to anticipate the type of normative systems that they will comply with. 
\end{enumerate}

The remainder of the paper is organised as follows. Section \ref{sec:background} provides the necessary background to understand EGT. Section \ref{sec:model} describes our framework, whereas Section \ref{sec:empirical} illustrates its empirical evaluation. Section \ref{sec:rw} reviews the state of the art in norm synthesis, and Section \ref{sec:conclusions} provides some concluding remarks and outlines possible future research. 
\section{Background: evolutionary game theory}
\label{sec:background}

We start by describing the framework of evolutionary game theory (EGT) \cite{smith1988evolution} and the key concept of \textit{evolutionarily stable strategy} (ESS). EGT combines population ecology with classical game theory. It considers a population of agents that repeatedly engage in strategic pairwise interactions by adopting different (pure) strategies. An ESS is a strategy that, if adopted by a majority of agents, no agent could benefit from using any alternative strategy \--- namely, the \textit{fitness} (i.e., the average payoff) of an agent using that strategy is higher than the fitness of any agent using alternative strategies. 

EGT provides a model of the underlying process whereby strategies change in a population. It assumes that successful strategies ``reproduce", growing in frequency with higher probabilities than less successful strategies. Evolutionarily stable strategies are attractor points of such a natural selection process whereby agents can converge to adopting an ESS. 

\begin{figure}[tb]
\begin{minipage}[t]{0.65\linewidth}
\centering
\includegraphics[width=1\linewidth]{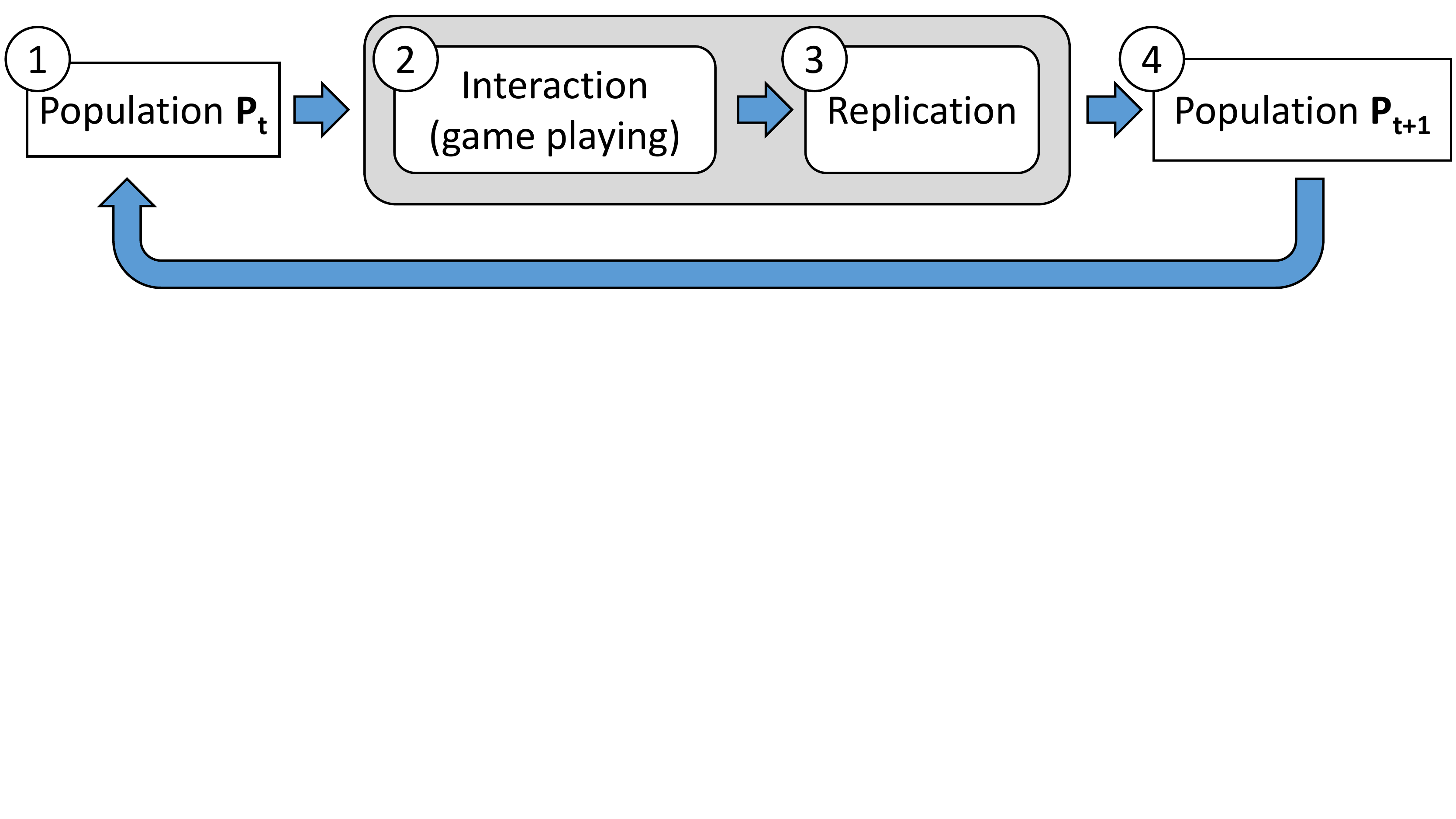}
\vspace{-1.175in}
\captionof{figure}{EGT model, composed of four phases: (1) a population $P_t$ is generated in which each mass of strategists has a certain size; (2) agents are randomly paired and play the game; (3) each mass of strategists grows in terms of their fitness when playing against different strategists; (4) a new population $P_{t+1}$ is generated in which each mass of strategists has grown in numbers proportional to its fitness.}
\label{fig:egtModel}
\end{minipage}
\hfill
\begin{minipage}[b]{0.31\linewidth}
\centering
\footnotesize
\begin{tabular}{cccll}
\cline{2-3}
\multicolumn{1}{c|}{} & \multicolumn{1}{c|}{\textbf{H}} & \multicolumn{1}{c|}{\textbf{D}} &  &  \\ \cline{1-3}
\multicolumn{1}{|c|}{\textbf{H}} & \multicolumn{1}{c|}{$V/2 - C/2$} & \multicolumn{1}{c|}{$V$} &  &  \\ \cline{1-3}
\multicolumn{1}{|c|}{\textbf{D}} & \multicolumn{1}{c|}{$0$} & \multicolumn{1}{c|}{$V/2$} &  &  \\ \cline{1-3}
\multicolumn{1}{l}{} & \multicolumn{1}{l}{} & \multicolumn{1}{l}{} &  & 
\end{tabular}
\captionof{table}{Payoff matrix of the Hawk-Dove game. When a Hawk (H) meets a Hawk, both get the resource half of the times with an injury cost. When a Hawk meets a Dove (D), the Hawk takes the resource and the Dove takes nothing. When a Dove meets a Dove, they share the resource.}
\label{tab:hawkdove}
\end{minipage}
\label{fig:egt}
\end{figure}

Figure \ref{fig:egtModel} graphically illustrates the EGT model. It considers an initial population of agents $P_t$ that adopt different strategies to play a game (Figure \ref{fig:egtModel}.1). First, agents are paired off randomly and play the game (Figure \ref{fig:egtModel}.2). Each strategy has a certain \textit{fitness} that quantifies the average payoff to an agent that adopts the strategy to play against other strategists. Strategies are then replicated (Figure \ref{fig:egtModel}.3), growing in frequency proportionally to their relative fitness with respect to the average fitness of the population. Then, a new population $P_{t+1}$ is generated that reflects the changes in strategy frequencies (Figure \ref{fig:egtModel}.4). Such population is then employed to repeat the process, which ends once the population remains stable between generations (that is, the frequencies of each strategy remain unchanged). 

Next, we detail the equations employed by EGT to perform strategy replication, also known as the \textit{replicator dynamics}. 

\subsection{Replicator dynamics}
\label{subsec:replicator}

We illustrate the replicator dynamics by making use of the first game analysed by Maynard Smith in \cite{smith1988evolution}: the \textit{Hawk-Dove} game. In this game, two individuals compete for some resource of value $V$. The players can adopt one of two different strategies:

\begin{itemize}
\item \textbf{Hawk (aggressive behaviour):} Fight to get the resource until either getting injured or the opponent backs down. 
\item \textbf{Dove (collaborative behaviour):} Back down if the opponent shows aggressive behaviour. Share the resource if the opponent shows collaborative behaviour.
\end{itemize}

The payoff matrix of this game is depicted in Table \ref{tab:hawkdove}. When a Hawk (H) meets a Hawk, both engage in conflict and have $50\%$ chance to get the resource and $50\%$ chance to get injured. Hence, both get payoff $V/2$ less an injury cost $C/2$. When a Hawk meets a Dove (D), the Hawk takes the whole resource and the Dove takes nothing (payoffs $V$ and $0$, respectively). When two Doves encounter, they equally split the resource (payoff $V/2$).

Consider a population of agents that adopt either strategy H or strategy D. Let $F(\mathit{H}) \in [0,1]$ be the frequency of Hawk strategists in the population, and $F(\mathit{D}) \in [0,1]$ the frequency of Dove strategists. Note that $F(\mathit{H}) + F(\mathit{D}) = 1$. Let us denote as $\rho(\mathit{H,D})$ the payoff to a Hawk when playing against a Dove, and analogously for other strategy pairs. We assume that each strategist has an initial fitness $f_0$. The fitness of each strategy will depend on: (1) the \textbf{payoff} to an agent when encountering either a Hawk or a Dove, and (2) the \textbf{probability} to encounter each one of these, which actually is a representation of the frequency of strategists of each type. Then, the fitness $f$ of each strategy can be computed as:
\begin{flalign}
f(\mathit{H}) & = f_0(\mathit{H}) + F(\mathit{H}) \cdot \rho(\mathit{H,H}) + F(\mathit{D}) \cdot \rho(\mathit{H,D}) \label{eq:fitnessHawk} \\
f(\mathit{D}) & = f_0(\mathit{D}) + F(\mathit{H}) \cdot \rho(\mathit{D,H}) + F(\mathit{D}) \cdot \rho(\mathit{D,D}) \label{eq:fitnessDove}
\end{flalign}
$f_0(\mathit{H})$ and $f_0(\mathit{D})$ being the initial fitness of Hawks and Doves, respectively. 

In this manner, the fitness of a Hawk is computed as the summation of its initial fitness, the probability of encountering a Hawk times the payoff to the Hawk when that happens, and the probability of encountering a Dove times the payoff to the Hawk when that happens. The fitness of a Dove is computed analogously. 

Agents reproduce in numbers proportional to their fitnesses. In the next generation, the frequency of Hawks and Doves is updated in terms of their relative fitnesses with respect to the average fitness of the Hawk-Dove population. Then, if Hawks perform better than average they will grow in frequency, and if Doves perform worse than average they will decrease in frequency. Formally, the frequencies of Hawks and Doves are updated as:
\begin{eqnarray}
F(\mathit{H})' = F(\mathit{H}) + F(\mathit{H}) \cdot \big[f(\mathit{H})-\theta \big] \label{eq:replicationHawk}\\
F(\mathit{D})' = F(\mathit{D}) + F(\mathit{D}) \cdot \big[f(\mathit{D})-\theta \big]
\label{eq:replicationDove}
\end{eqnarray}
where $\theta$ is the weighted average fitness of the Hawk-Dove population, computed as:
\begin{equation}
\theta = F(\mathit{H}) \cdot f(\mathit{H}) + F(\mathit{D}) \cdot f(\mathit{D})
\label{eq:avgFitnessEgt}
\end{equation}

In biology, replication models the natural process whereby the fittest individuals are more likely to survive and to reproduce than less fit ones. In economic settings (such as multi-agent systems), replication provides a model of imitation \cite{iwai1984schumpeterian,bjornerstedt1994imitation} whereby the agents tend to imitate strategists that appear to perform well, i.e., have a higher fitness, thereby adopting their strategies over time. Then, if a strategy is fitter than the average, agents will be more likely to adopt it than to adopt a less fit one. 

As previously mentioned, the replication process can eventually lead the population to a point of equilibrium in which the frequencies of each strategy do not change over time because their fitnesses are equal. When this happens, the population can be either \textit{monomorphic} (a majority of agents adopt the same strategy) or \textit{polymorphic} (the agents adopt a variety of strategies). If the population composition can be restored after a disturbance\footnote{Provided that the disturbance is not too large. For example, a small number of mutant strategists joins the scenario, and after some time they are ``eliminated" by dominant strategists.}, then it is said that the population is in an \textit{evolutionarily stable state}. If such population is monomorphic, then it is said that the strategy adopted by its agents is an ESS. Next, we detail the conditions for a strategy to be evolutionarily stable. 

\subsection{Evolutionarily stable strategies}
\label{subsec:ess}

A strategy is evolutionarily stable if, once a majority of agents adopt it, their fitness is higher than that of any possible mutant strategist. Otherwise, agents may be tempted to switch to alternative strategies, and the strategy would be unstable. As an example, consider a population composed mainly of Hawks with a small proportion of Doves. That is, $F(\mathit{H}) \simeq 1$ and $F(\mathit{D}) \ll 1$. If Hawk is an evolutionarily stable strategy, then it must satisfy that either:
\begin{enumerate}
\item \textbf{Hawk is a best response to itself}. That is, Hawks must perform better than Doves when playing against Hawks. 

\vspace{0.05in}
or
\vspace{0.05in}

\item \textbf{Hawk is a best response to Dove}. In other words, Hawk is not necessarily a best response to itself, but Hawks must perform better against Doves than they perform against themselves.
\end{enumerate}
Formally, this amounts to satisfying either condition \ref{eq:ess1} or condition \ref{eq:ess2} below. 
\begin{flalign}
\rho(\mathit{H,H}) & > \rho(\mathit{D,H}) \label{eq:ess1} \\ 
\rho(\mathit{H,H}) & = \rho(\mathit{D,H})\mbox{ \, and \, }  \rho(\mathit{H,D})  > \rho(\mathit{D,D}) \label{eq:ess2}
\end{flalign}
It is obvious that Dove is not an ESS, since a population of Doves can be invaded by a Hawk mutant. The only evolutionarily strategy is Hawk, as long as the injury cost $C$ is lower than the value of the resource $V$ (so that it is worth getting injured in order to obtain the resource). If the injury cost is greater than the value of the resource, then there is no ESS. 

%

\section{Evolutionary norm synthesis}
\label{sec:model}

In this section we introduce our framework for the synthesis of \textit{evolutionarily stable normative systems} for non-deterministic settings \--- hereafter, referred to as our ``evolutionary norm synthesis system", or \acro{ENSS} for shorter. In Section \ref{subsec:outline}, we start by providing a general overview of the \acro{ENSS} operation. Then, we provide some basic definitions and formally define our problem in Section \ref{subsec:problem}. Finally, we describe in detail how the \acro{ENSS} performs evolutionary norm synthesis in Section \ref{subsec:model}.

\subsection{Evolutionary norm synthesis system (ENSS)}
\label{subsec:outline}

Our framework is intended to synthesise norms that achieve coordination in a MAS and are evolutionarily stable. With this aim, we will assume that our system considers no previous knowledge about the potential interactions of the agents, neither about their outcomes. Instead, it will learn these situations from the observation of the agents' activities at runtime, and will synthesise norms to coordinate them. Likewise EGT (Section \ref{sec:background}), our system will endow each norm with: (i) a \textit{fitness} value that quantifies its utility to coordinate the agents in a coordination setting; and (ii) a \textit{frequency} value that stands for the proportion of agents that have adopted the norm. The \acro{ENSS} enacts an evolutionary process whereby the fittest norms prosper and increase in frequency, and the less fit ones are ultimately eliminated. Eventually, the agents converge to adopting a set of norms that are evolutionarily stable. 

\begin{figure}
\centering
\includegraphics[width=0.7\linewidth]{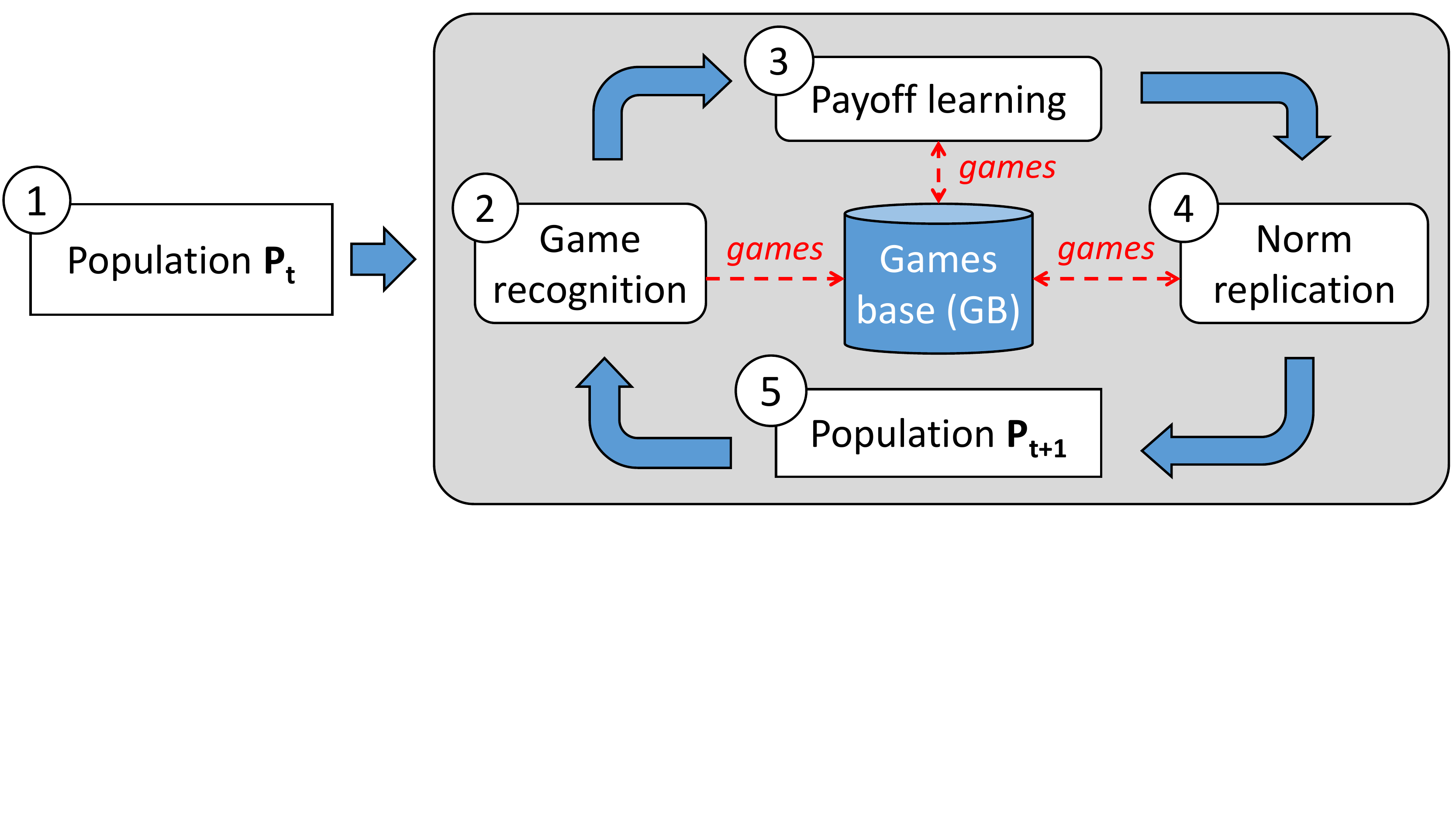}
\vspace{-0.6in}
\caption{Graphical representation of our evolutionary norm synthesis system (\acro{ENSS}). It starts in (1), with an initial population of agents $P_t$; (2) agents interact for a given period of time, during which the \acro{ENSS} identifies the coordination situations (games), and creates their corresponding norms; (3) the \acro{ENSS} empirically computes norms' payoffs; (4) norms are replicated, growing in numbers proportional to their fitness; (5) a new population $P_{t+1}$ is generated in which the size of the set of agents adopting each norm is proportional to its frequency. Such new population is employed to perform a new iteration of the process.}
\label{fig:neModel}
\end{figure}

Figure \ref{fig:neModel} illustrates the evolutionary process implemented by our \acro{ENSS}. It starts with an initial agent population $P_t$ whose agents employ no norms to coordinate. Then, it repeatedly performs the following tasks: 
\begin{itemize}
\item \textbf{Game recognition.} The \acro{ENSS} identifies new situations that require coordination by observing agents' interactions, and keeps track of them as games in a \textit{Games Base} (GB). For each recognised game, the \acro{ENSS} creates norms prescribing different coordination strategies, and sends each norm to different agents. The result will be a heterogeneous population whose agents use different strategies to coordinate. 

\vspace{0.05in}

\item \textbf{Payoff learning.} Our \acro{ENSS} continuously monitors the agents' game play in order to cumulate evidence about the performance (the payoffs) of each norm in terms of the frequency with which it successfully coordinates the agents.

\vspace{0.05in}

\item \textbf{Norm replication.} The \acro{ENSS} computes norms' fitnesses based on their average payoffs up to a given time, and replicates norms in numbers proportional to their fitness: the frequency of those norms fitter than average will increase, while that of those norms less fit than average will decrease. The output will be a new population $P_{t+1}$ in which the size of the set of agents adopting each norm is proportional to its frequency. 
\end{itemize}

The \acro{ENSS} will repeatedly perform these tasks until either it converges, or the MAS stops running. We say that the \acro{ENSS} has converged to a stable population once the frequency of each norm remains unchanged from population $P_t$ to population $P_{t+1}$ for a given number of iterations $I$. Upon convergence, if a large majority of agents have adopted the same normative system, then we say that their normative system is evolutionarily stable.

\begin{figure}
\centering
\includegraphics[width=1\linewidth]{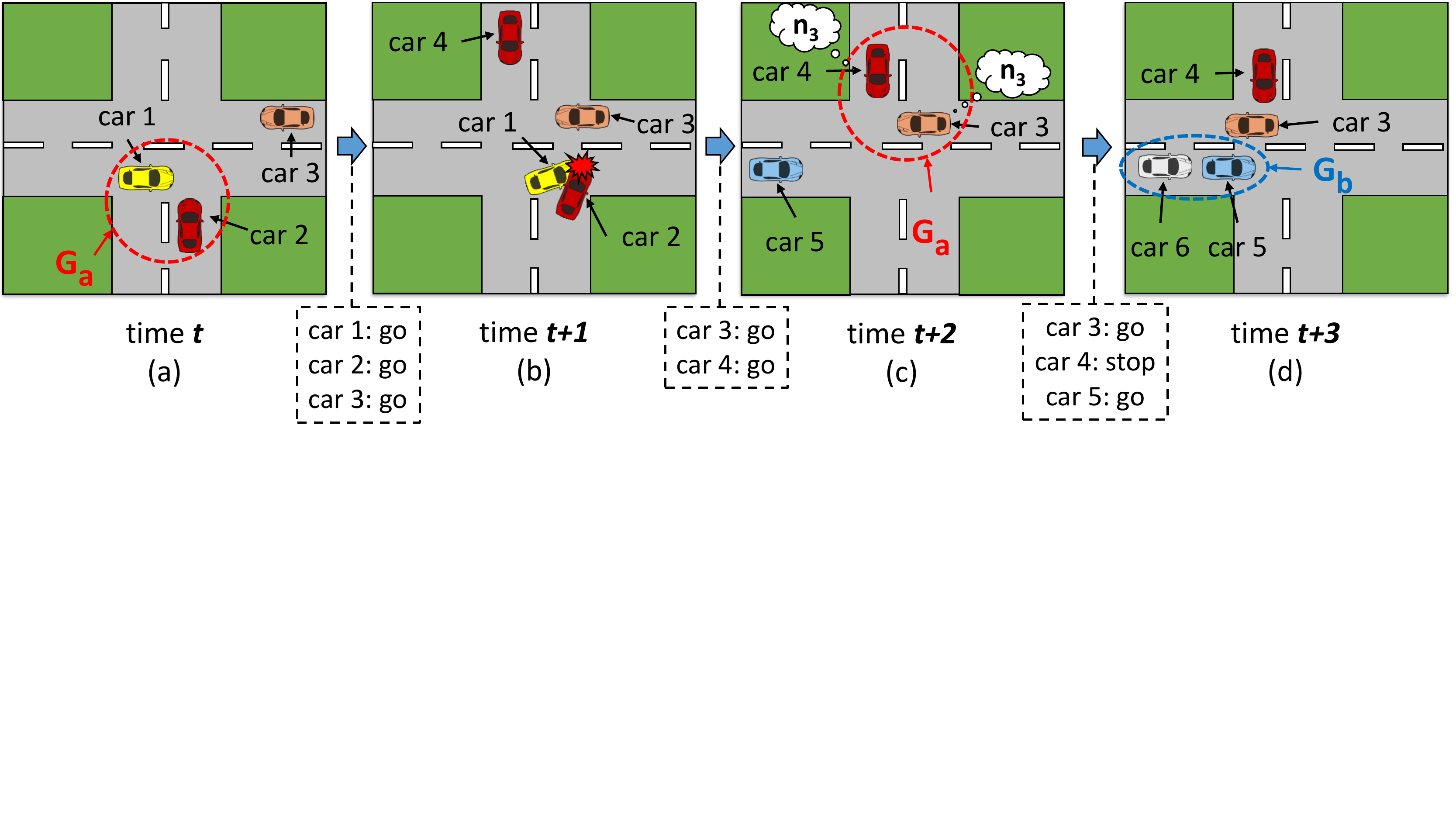}
\vspace{-1.4in}
\captionof{figure}{(a) A junction at time $t$, with cars 1 and 2 playing a $2$-role game; (b) At time $t+1$, cars 1 and 2 collide after both perform action \textit{``go"}; (c) At time $t+2$, cars 3 and 4 play again the same game, this time by having norms to coordinate; (d) At time $t+3$, cars 3 and 4 have avoided a collision. Also, car 6 plays a $1$-role game in which it has to decide whether to go forward, or to stop to keep a security distance with car 5.}
\label{fig:games}
\end{figure}

\begin{figure}
\begin{minipage}[t]{0.23\linewidth}
\centering
\includegraphics[width=1.35\linewidth]{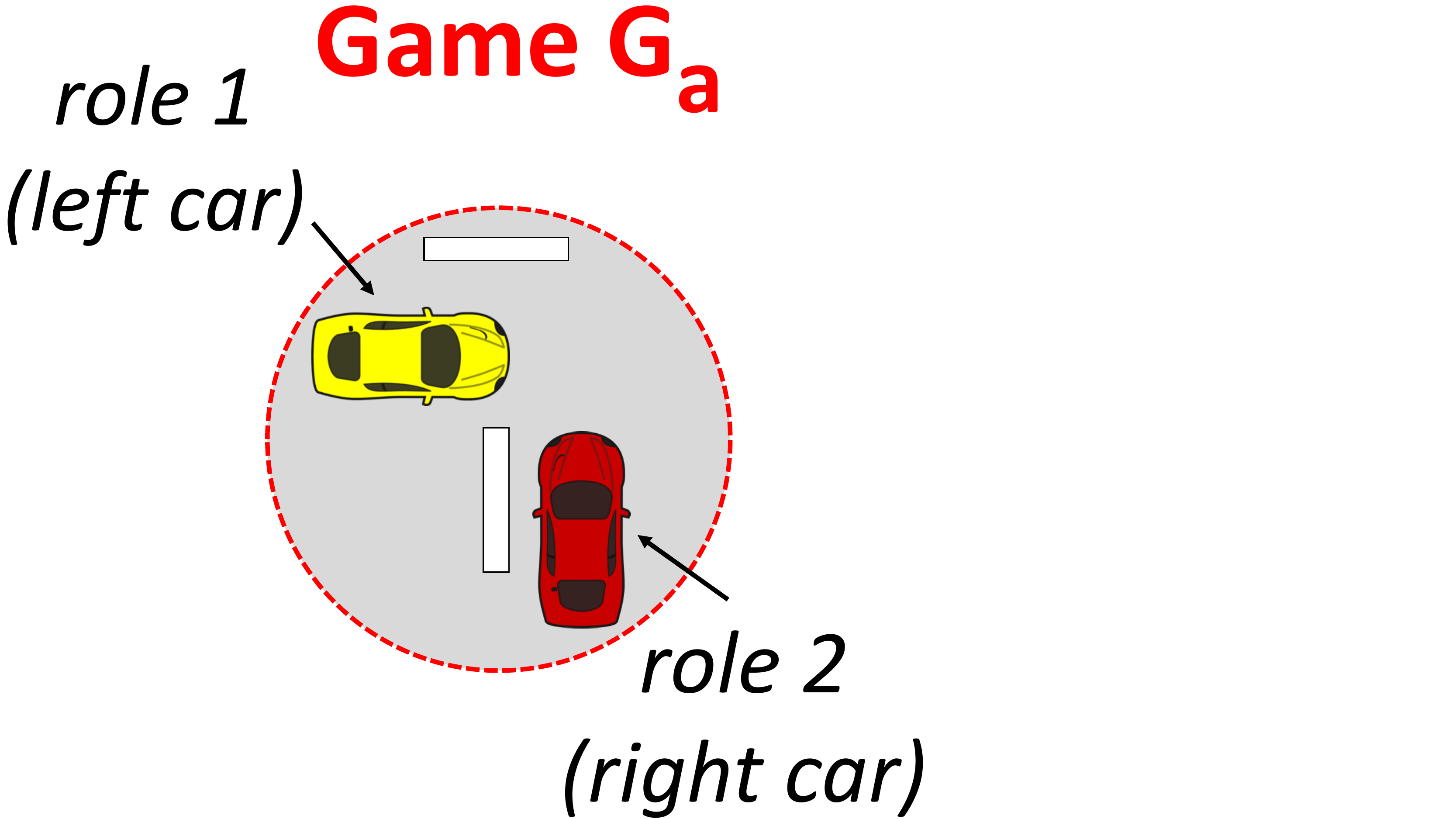}
\vspace{-.1in}
\captionof{figure}{$2$-role game played by cars 1 and 2 in Figure \ref{fig:games}a.}
\label{fig:game1}
\end{minipage}
\hfill
\begin{minipage}{0.39\linewidth}
\centering
\footnotesize
\begin{tabular}{cccc}
\multicolumn{1}{l}{} & \multicolumn{1}{l}{} & \multicolumn{2}{c}{\textbf{role 2}} \\
\multicolumn{1}{l}{} & \multicolumn{1}{l}{} & \multicolumn{2}{c}{\textit{\textbf{(right car)}}} \\ \cline{3-4} 
\multirow{3}{*}{\textbf{\begin{tabular}[c]{@{}c@{}}role 1\\ \textit{(left car)}\end{tabular}}} & \multicolumn{1}{c|}{} & \multicolumn{1}{c|}{\textit{go}} & \multicolumn{1}{c|}{\textit{stop}} \\ \cline{2-4} 
 & \multicolumn{1}{|c|}{\textit{go}} & \multicolumn{1}{c|}{0,0} & \multicolumn{1}{c|}{1,0.7} \\ \cline{2-4} 
 & \multicolumn{1}{|c|}{\textit{stop}} & \multicolumn{1}{c|}{0.7,1} & \multicolumn{1}{c|}{0.7,0.7} \\ \cline{2-4} 
\end{tabular}
\vspace{0.28in}
\captionof{table}{Rewards for a given role in game $G_a$ (depicted in Figure \ref{fig:game1}) given each possible outcome of the game.}
\label{tab:game}
\end{minipage}
\hfill
\begin{minipage}{0.3\linewidth}
\centering
\begin{tabular}{ccc}
 & \multicolumn{2}{c}{\textbf{Prohibitions}} \\ \cline{2-3} 
\multicolumn{1}{c|}{\textbf{Norm}} & \multicolumn{1}{c|}{\textit{role 1}} & \multicolumn{1}{c|}{\textit{role 2}} \\ \hline
\multicolumn{1}{|c|}{$n_1$} & \multicolumn{1}{c|}{---} & \multicolumn{1}{c|}{---} \\ \hline
\multicolumn{1}{|c|}{$n_2$} & \multicolumn{1}{c|}{go} & \multicolumn{1}{c|}{---} \\ \hline
\multicolumn{1}{|c|}{$n_3$} & \multicolumn{1}{c|}{---} & \multicolumn{1}{c|}{go} \\ \hline
\multicolumn{1}{|c|}{$n_4$} & \multicolumn{1}{c|}{go} & \multicolumn{1}{c|}{go} \\ \hline
\end{tabular}
\vspace{0.1in}
\captionof{table}{Different norms to coordinate any two cars that play game $G_a$ in Figure \ref{fig:game1}.}
\label{tab:norms}
\end{minipage}
\label{fig:game}
\end{figure}

Let us illustrate the operation of our \acro{ENSS} with an example. Consider a traffic scenario where agents are cars, and the coordination task is to ensure that cars reach their destinations as soon as possible without colliding. The actions available to the cars are \textit{``go"} forward and \textit{``stop"}. Figure \ref{fig:games}a depicts a junction at time $t$ with three cars. Of these, cars 1 and 2 (circled in red) require coordination in order to avoid colliding. Figure \ref{fig:games}b illustrates a collision between these cars at time $t+1$ after both have performed action \textit{``go"}. The \acro{ENSS} will detect such a coordination situation, and will model it as a \textit{one-shot} game (labelled as $G_a$ in Figure \ref{fig:game1}) with two \textit{roles}: a car on the left (role 1) and a car on the right (role 2), both perceiving each other. The \acro{ENSS} will create the following norms for $G_a$, which are listed in Table \ref{tab:norms}: 

\begin{itemize}
\item Norm $n_1$ establishes no prohibitions, and hence a car is free to go forward (not giving way) when coming from either the left or the right (no matter the role it plays). 

\item Norm $n_2$ says that a car is prohibited to go forward when coming from the left (when playing role 1). In practice, $n_2$ stands for a ``give way to the right" norm. Analogously, norm $n_3$ stands for a ``give way to the left" norm. 

\item Norm $n_4$ stands for a ``give way always" norm, i.e., it prohibits a car to go forward once it is playing either role 1 or role 2. 
\end{itemize}

The \acro{ENSS} will deliver each one of these norms to different cars. For instance, it will deliver norm $n_1$ to 25\% of the cars, and the same applies to norms $n_2, n_3$ and $n_4$. Thereafter, the \acro{ENSS} will detect once the cars play game $G_a$, and will monitor their outcome in order to compute the payoffs of each norm. For instance, at time $t+2$ (Figure \ref{fig:games}c) two new cars play $G_a$ (car 3 plays role 1, and car 4 plays role 2). Suppose that both cars have and apply norm $n_3$ (represented as a thought bubble containing $n_3$). Thus, at time $t+3$ (Figure \ref{fig:games}d) car 3 goes forward and car 4 stops, avoiding collisions. The \acro{ENSS} will monitor this positive outcome, increasing the payoff of norm $n_3$ whenever two cars jointly use it to coordinate. 

Also, at time $t+3$ the \acro{ENSS} will detect a new one-role game played by car 6 (labelled as $G_b$), which can stop, keeping a security distance with car 5, or to go forward, possibly colliding with car 5 in case it stops. The \acro{ENSS} will create new norms for this game, and will deliver them to the cars. In this way, each car will incrementally build a personal set of norms aimed to coordinate in different games.


\subsection{Basic definitions and problem statement}
\label{subsec:problem}

We model a multi-agent system (MAS) as a state transition system with a finite set of agents $Ag$, and a finite set of actions $Ac$ available to these agents. Let $S$ be the set of the states of the system. We consider a language $\mathcal{L}$ (with an entailment relation $\models$) to describe what is true and false in each state. We adopt a synchronous model in which agents interact in some state, perform a collection of actions, and lead the MAS from its previous state to a new one. 

In each MAS state, agents may engage in strategic interactions in which they need coordination in order to avoid undesirable outcomes. Hereafter, we will refer to undesirable outcomes as \textit{conflicts}. We will model coordination situations as \textit{one-shot} games. Each game will have a \textit{context} that describes the situation of each agent playing the game before acting. Each one of these agents will take on one \textit{role}, which defines the part she plays in the game, and the actions she can perform. Hereafter, we will refer to an agent taking on role $i$ as \textit{player} $i$. Once the players of a game select a joint action, each player receives a \textit{reward} that quantifies to what extent she has successfully avoided conflicts. 

\begin{mydef}[$m$-role game]
\label{def:game}
An $m$-role game is a tuple $G=\langle \varphi, R, A, \mathcal{R}\rangle$, where:
\begin{itemize}
\item $\varphi$ is an expression of $\mathcal{L}$ describing the starting conditions (context) of the game.
\item $R=\{1, \ldots, m\}$ is a set of $m$ agent roles, one per agent playing the game.
\item $A=\langle A_1, \ldots, A_m \rangle$ is an $m$-tuple of action sets available to each role, where $A_i \subseteq Ac$ is the set of actions available to the agent taking on role $i$.
\item $\mathcal{R}=\langle r_1, \ldots, r_m \rangle$ is an $m$-tuple of reward functions of the form $r_i: \prod_{i=1}^m A_i \rightarrow \mathbb{R}_{\geq 0}$, each one returning a reward to player $i$ after all players draw a joint action $\vec{a}$.\footnote{$\vec{a}=\langle a_1, \ldots, a_m \rangle$ stands for a vector of actions, one for each player.}
\end{itemize}
\end{mydef}

For example, we will define game $G_a$ in Figure \ref{fig:game1} as a $2$-role game where role 1 represents a car on the left, and role 2 a car on the right. Both roles have available actions \textit{``go"} and \textit{``stop"}. This game will be a tuple $G_a=\langle \varphi, R, A, \mathcal{R}\rangle$, where its context $\varphi$ can be formally interpreted as: \textit{``there is a car playing role 1 that perceives a car playing role 2 on its right, and vice versa"}, $R=\{1,2\}$ is the set of roles, $A=\langle\{go,stop\},\{go,stop\}\rangle$ is the set of action spaces of each role, and $\mathcal{R}$ contains the reward functions of Table \ref{tab:game}. 
Specifically, two cars that jointly go forward will collide and get reward 0 each. Once at least one of the cars stops, both will avoid colliding. In that case, a car will get reward 1 if it is able to progress (to move forward), and reward 0.7 otherwise. 

In general, in each state of the MAS the same $m$-role game can be simultaneously played by different groups of agents. We assume that each agent playing an $m$-role game in a given MAS state has available a function $\mathit{role}$ that returns the role she plays in that game. 

Given an $m$-role game, a norm stands for a coordination strategy that specifies what an agent is prohibited to do when playing each possible role. For example, each norm in Table \ref{tab:norms} establishes different prohibitions for any car playing either role 1 or 2 in game $G_a$. Formally, a norm is a (possibly empty) set of constraints that restricts the action space of each role of a game by prohibiting certain actions.

\begin{mydef}[Norm]
\label{def:norm}
Given an $m$-role game $G=\langle \varphi, R, A, \mathcal{R}\rangle$, a norm to coordinate the agents in $G$ is a pair $\langle \psi, prh \rangle$, where: 
\begin{itemize}
\item $\psi \in \mathcal{L}$ is the precondition of the norm; and
\item $prh:R \rightarrow 2^{Ac}$ is a function that returns the set of actions that an agent is prohibited to perform when taking on role $i$, where $prh(i) \in 2^{A_i}$ for all $i \in R$.
\end{itemize}
\end{mydef}
Let $G=\langle \varphi, R, A, \mathcal{R}\rangle$ be an $m$-role game, and $n = \langle \psi, prh \rangle$ a norm. We say that $n$ applies in $G$ if the precondition of $n$ satisfies the context of $G$, namely if $\varphi \models \psi$. Hereafter, we will refer as the set of norms that apply in a game $G$ as the \textit{norm space} of the game, and we will denote it by $N_G$. For instance, the norm space of game $G_a$ (Figure \ref{fig:game1}) is $N_{G_a}=\{n_1, n_2, n_3, n_4\}$ (see Table \ref{tab:norms}). 

Agents in a MAS may play multiple, different $m$-role games. Henceforth, we shall denote the set of games that agents can play as $\mathcal{G} = \{G_1,\ldots,G_s\}$. A normative system is a set of $|\mathcal{G}|$ norms  that provides an agent with the means to coordinate in each possible game in $\mathcal{G}$. Following our example, each car will have one norm out of the norm space $N_{G_a}$ to coordinate in $G_a$, one norm out of $N_{G_b}$ to coordinate in $G_b$, and so on for each game. 

\begin{mydef}[Normative system]
Let $\mathcal{G}$ be a set of $m$-role games. A normative system is a set of norms $\Omega$ such that for each game $G \in \mathcal{G}$ there is one and only one norm $n \in \Omega$ and $n \in N_G$.
\label{def:ns}
\end{mydef}

First of all, each agent in a MAS $ag_i \in Ag$ counts on her own normative system $\Omega_i$. Thus, in general we assume that a MAS is composed of a heterogeneous population whose agents may have different normative systems. 

Let $Ag' \subseteq Ag$ be a group of agents playing an $m$-role game $G=\langle \varphi, R, A, \mathcal{R}\rangle$ at a given time $t$. Each agent counts on one and only one norm out of her normative system that applies in the game and prohibits her to perform some actions. We consider: 
\begin{itemize}
\item an injective function $\pi: R \rightarrow Ag$ that maps each role in $R$ to one agent in $Ag'$, namely to the agent enacting that role in $G$ at time $t$; and
\item a function $\eta:Ag \times \mathcal{G} \rightarrow N_{G_1}\cup\ldots\cup N_{G_s}$ that given an agent $ag_i$ and a game $G$, tells us the norm in the normative system of the agent, $\Omega_i$, to apply in $G$. 
\end{itemize}

This allows us to define $\vec{n}=\langle \eta(\pi(1),G), \ldots, \eta(\pi(m),G) \rangle$ as the combination of norms that the normative systems of the agents in $Ag'$ prescribe them to apply in $G$ at time $t$. Notice that $\eta(\pi(i),G)$ stands for the norm for game $G$ in the normative system of the agent playing role $i$. We assume that the agents always comply with the prohibitions prescribed by their norms. Therefore, based on the norms in $\vec{n}$, the agents in the game will perform a tuple of actions $\vec{a}=\langle a_1,\ldots,a_m \rangle$, where $a_i$ is an action that is not prohibited by norm $\eta(\pi(i),G)$ for role $i$.\footnote{In principle, given a combination of norms $\vec{n}$ applicable to a group of agents, it is not possible to assume beforehand the joint action $\vec{a}$ that these agents will perform. However, we assume that the actions in $\vec{a}$ will comply with the prohibitions established by their respective norms in $\vec{n}$.} After the agents perform a joint action $\vec{a}$, each player $i$ obtains a reward $r_i(\vec{a})$. 

Let us illustrate these definitions with cars 3 and 4 in Figure \ref{fig:games}c, which play game $G_a$ at time $t+2$ by enacting roles 1 and 2, respectively. Say that these cars have normative systems $\Omega_3$ and $\Omega_4$, respectively, and that both normative systems have $n_3$ as the applicable norm in game $G_a$. That is, $\eta(3,G) = n_3$ and $\eta(4,G) = n_3$. Thus, these cars will play with norm combination $\vec{n}=\langle n_3, n_3 \rangle$, which means that car 4 will be prohibited to go forward (and hence will stop). In practice, these cars will perform a joint action $\vec{a}=\langle go,stop \rangle$. Then, the reward to each role $i \in \{1,2\}$ at time $t+3$ can be computed as $r_i(\langle go,stop \rangle)$. 

Note therefore that the reward that an agent can expect obtain from a game depends on the norm that she uses to play the game, the role that she plays, and the norms that the other players use. Thus, at a given time an agent may get a high reward once she plays against agents with a particular combination of applicable norms, and at a different time she may get a low reward when playing against agents with a different norm combination. 

In practice, given an $m$-role game and the norms that apply in it, the agents will play an infinitely repeated one-shot game of norms against norms, namely a \textit{norms game}, in which the norms used by the agents to play the game over time will lead them to obtain a history of rewards. Thus, a norms game will consist of an $m$-role game, the norm space of the game, and a \textit{memory} that contains the history of rewards obtained by the agents once they repeatedly play the $m$-role game over time. 

\begin{mydef}[Norms game]
\label{def:game}
A norms game is a tuple $NG=\langle G, N_G, H, P\rangle$, where:
\begin{itemize}
\item $G=\langle \varphi, R, A, \mathcal{R} \rangle$ is an $m$-role game.
\item $N_G$ is the norm space of $G$, namely the set of norms that apply in $G$.
\item $H=\langle h_0,\ldots, h_\omega \rangle$ is the memory of the game over a time window $[t_0,t_{\omega}]$, where $h_j=\langle \vec{n}^j, \vec{r}^j \rangle$ such that $\vec{n}^j$ is the norm combination used by the agents to play $G$ at time $t_j$, and $\vec{r}^j$ is the vector of rewards they obtained at time $t_j$ (one for each player). 
\item $P=\langle \rho_1, \ldots, \rho_m \rangle$ is an $m$-tuple of payoff functions of the form $\rho_i: \prod_{i=1}^{|R|} N_G \times H \rightarrow \mathbb{R}_{\geq 0}$, which return the expected payoff to player $i$ based on the memory of the game $H$ and a combination of norms $\vec{n} \in N_G^{|R|}$ applicable to all players. 
\end{itemize}
\end{mydef}

Intuitively, the expected payoff of a combination of norms $\vec{n}$ tells us how successful that norm combination has been historically to coordinate the players of the game. Such a payoff is computed based on the rewards obtained by the players of the game within a time window. Further on, we provide an equation to compute the expected payoff in Section \ref{subsec:payoffs}.  

At this point we import from EGT the concept of \textit{fitness} introduced in Section \ref{sec:background}. Thus, given a norms game, the fitness of each one of its norms quantifies the \textit{average payoff} that an agent can expect to obtain when using the norm to play the game by enacting different roles and by playing against agents with different norms. Formally:

\begin{mydef}[Norm fitness]
Given a norms game $NG=\langle G, N_G, H, P\rangle$, the fitness of a norm $n \in N_G$ is represented as $f(n,NG, t) \in \mathbb{R}$, where $f$ stands for the norm fitness function at a particular point in time $t \in \mathbb{N}$.
\label{def:fitness}
\end{mydef}

Now we are ready to introduce the problem that we address in this paper. Let us assume a population of rational agents that will tend to adopt fitter norms. Given a norms game $NG = \langle G, N_G, H, P\rangle$, our aim is to find a norm $n \in N_G$ such that, once it is used by all the agents to play $G$, there is no agent that can derive a greater fitness by using any alternative norm $n' \in N_G$. Then, the rational choice for all the agents will be complying with norm $n$. In terms of EGT, this amounts to saying that norm $n$ is \textit{evolutionarily stable} (Section \ref{subsec:stability}), since no agent could be ever tempted to use alternative norms to play the game. Thus, given a norms game, our aim is to find a norm $n$ applicable in the game such that $n$ is evolutionarily stable. Likewise, given a collection of norms games, we aim to find a normative system that contains one evolutionarily stable norm for each norms game. Formally, our research problem is as follows. 

\begin{mydef}[Norm synthesis problem]
\label{def:problem}
Given a set of agents $Ag$ and a set of norms games $\mathcal{NG}$, our aim is to find a normative system $\Omega$ such that, from some time $t_u$ onwards, the following conditions hold:
\begin{enumerate}
\item \textbf{All agents adopt $\Omega$}. That is, $\Omega_i = \Omega$ for each agent $ag_i \in Ag$.
\item \textbf{There is no norm in the normative system whose fitness is outperformed by that of an alternative norm}. Namely, there is no norms game $NG=\langle G, N_G, H, P \rangle \in \mathcal{NG}$ and norm $n \in \Omega$, $n \in N_G$, such that $f(n',NG, t) > f(n,NG, t)$ for some alternative norm $n' \in N_G$ and time $t \geq t_u$.
\end{enumerate} 
\end{mydef}


\subsection{Formal model for evolutionary norm synthesis}
\label{subsec:model}

We now describe the tasks that our \acro{ENSS} performs to synthesise a normative system that solves  the norm synthesis problem in Definition \ref{def:problem}. That is, \textit{game recognition} (labelled as 2 in Figure \ref{fig:neModel}), \textit{payoff learning} (labelled as 3) and \textit{norm replication} (labelled as 4). These tasks allow the \acro{ENSS} to detect and abstract as games the coordination situations that the agents might encounter, and to carry out the evolutionary process whereby the agents can ultimately adopt an evolutionarily stable normative system. 

\subsubsection{Recognising new games from observation}
\label{subsubsec:games}

Game recognition is achieved by observation of the agents' activities. Agents interact in an environment for a given number of time steps, and the \acro{ENSS} monitors their activities at regular time intervals. At each time step, the \acro{ENSS} tries to detect new, untracked coordination situations that it abstracts as new $m$-role games. With this aim, we take inspiration from the work in \cite{morales2013automated,morales2014minimality,morales2015liberality}, which performs automatic detection of coordination situations by detecting agent interactions that lead to undesirable outcomes (conflicts). Analogously, we consider that any type of $m$-agent interaction that may lead to conflicts is an $m$-role game. 

As in \cite{morales2013automated,morales2014minimality,morales2015liberality}, we assume that conflicts can be detected at runtime, and that the agents involved in a conflict are the ones responsible for the conflict. Moreover, we assume that a conflict at time $t$ is caused by the actions that the agents performed at time $t-1$. At time $t$, the \acro{ENSS} builds a new $m$-role game following the next steps:
 
\begin{enumerate}
\item \emph{Detect a new conflict at time $t$.} Notice that detecting conflicts requires the use of domain-dependent knowledge to retrieve the conflicts at a given time. Examples of conflicts are: collisions in a traffic scenario.  
\item \emph{Describe the situation involving the $m$ conflicting agents.} This amounts to generating an expression $\varphi \in \mathcal{L}$ that describes the situation involving these agents in the state prior to the occurrence of the conflict (i.e., at time $t-1$). 
\item \emph{Create a new $m$-role game to model coordination} as $G=\langle \varphi, R, A, \mathcal{R}\rangle$ with expression $\varphi$ as its context, $R$ as the set of roles played by the agents at time $t-1$, $A$ as the set of the actions available to these agents at time $t-1$, and $\mathcal{R}$ as the set of reward functions.\footnote{Note that assessing the rewards for a given game requires the use of domain-dependent knowledge.}
\end{enumerate}

Then, if the new $m$-role game $G$ does not exist in the Games Base (see Figure \ref{fig:neModel}), the \acro{ENSS} will add it in order to be able to detect when the agents play the game. 

We already illustrated this process in Section \ref{subsec:problem} by creating game $G_a=\langle \varphi, R, A, \mathcal{R} \rangle$ from the interaction of cars 1 and 2 at time $t$ (Figure \ref{fig:games}a). $G_a$ has $\varphi \in \mathcal{L}$ as its context, $R=\{1,2\}$, $A=\{\langle go,stop\rangle, \langle go,stop\rangle\}$, and $\mathcal{R}$ as its set of reward functions (Table \ref{tab:game}).


After creating a new $m$-role game $G$, the \acro{ENSS} will create its norm space $N_G$ by:

\begin{enumerate}
\item identifying the actions performed by the $m$ conflicting agents in the transition from the state prior to the conflict to the state containing the conflict, i.e., from time $t-1$ to $t$. 

\item creating norms prohibiting different roles to perform the conflicting actions in the game, each  having $\varphi$ as its precondition, and a combination of prohibitions as its postcondition.
\end{enumerate}

After that, the \acro{ENSS} will create the corresponding norms game $NG=\langle G, N_G, H, P\rangle$, with the new $m$-role game $G$, $N_G$ as its norm space, $H$ as its tuple of history functions\footnote{Initially, each history function $h_i \in H$ will return an empty sequence of rewards. Consequently, each empirical payoff function $\rho_i \in P$ will return an undefined value.}, and $P$ as its set of expected payoff functions. Next, the norms of $N_G$ are uniformly distributed among the agents in a MAS. This guarantees that the normative systems in an agent population are heterogeneous, namely the agents will play game $NG$ by using different norms, and hence the \acro{ENSS} will be able to evaluate which ones do better.

Going back to the example of game $G_a$, the \acro{ENSS} will now create its norm space $N_{G_a}$ by first identifying action ``go" as the action performed by the conflicting cars 1 and 2 during the transition from time $t$ to time $t+1$ in Figure \ref{fig:games}. Then, it will create norms to prohibit to perform action ``go" to: none of the roles ($n_1$ in Table \ref{tab:norms}), role 1 ($n_2$), role 2 ($n_3$), and both roles ($n_4$). The \acro{ENSS} will now create and track the corresponding norms game $NG_a=\langle G_a, N_{G_a}, H, P \rangle$. Thereafter, it will deliver each norm to 25\% of the agents.

\subsubsection{Computing norms' payoffs empirically}
\label{subsec:payoffs}

The \acro{ENSS} continuously monitors the game play of the agents, detecting when they play each norms game, and keeping track of their rewards in the memory of the game. The \acro{ENSS} will exploit this knowledge in order to approximate the expected payoffs of the game at a given point in time based on the following principles:
\begin{enumerate}
\item Those norms that have allowed the agents obtain high rewards in the past can be expected to yield high rewards in future game plays. 
\item The rewards obtained by the agents more recently in the past are more valuable and informative for the payoff computation than older rewards. 
\end{enumerate}
Thus, the payoffs of a norms game are computed as follows. Let $NG=\langle G, N_G, H, P \rangle$ be a norms game, and $\vec{n}$ a combination of norms applicable to a group of agents playing $NG$ at a given time $t_\omega$. First, we will retrieve from $H$ a tuple $\langle \vec{r}^1, \ldots, \vec{r}^k \rangle$ with the rewards obtained by the agents in the $k$ times they played $NG$ with norm combination $\vec{n}$ within a time window $[t_0, t_\omega]$. Then, we will compute the expected payoff to player $i$ as the \textit{discounted average reward} to role $i$ within this time window:
\begin{equation}
\rho_i(\vec{n}, H) = \underbrace{\dfrac{1}{\sum\limits_{j=1}^k \beta^{k-j}}}_{\mathit{normalisation}} \ \underbrace{\sum\limits_{j=1}^k \vec{r}_i^j \cdot \beta^{k-j}}_{\mathit{discounted \ reward}}
\label{eq:payoff}
\end{equation}
where $\vec{r}_i^j$ is the $j$-th reward obtained by player $i$ in $NG$ within a time window $[t_0, t_\omega]$; and $\beta \in [0,1]$ is a discount factor. Intuitively, the right part of equation \ref{eq:payoff} computes the weighted summation of the rewards to player $i$ within a time window, where the $j$-th reward (an older reward) has a lower weight (is more discounted) than the $(j+1)$-th reward (a more recent one). The left part of the equation normalises the payoff by dividing the weighted summation by the summation of weights. 

At this point we recall that our system assumes a non-deterministic setting in which the outcomes of a game, and hence its rewards, cannot be assumed a priori. Instead, the \acro{ENSS} will assess the rewards for the players of a game by monitoring their outcomes a posteriori, i.e., once they have played the game. Thus, given an $m$-role game $G=\langle \varphi, R, A, \mathcal{R} \rangle$ played by a group of agents at a given time $t$, the \acro{ENSS} will approximate the reward function $r_i \in \mathcal{R}$ for player $i$ as an empirical reward function $\tilde{r}_i: \prod_{i=1}^{|R|}A_i \times \mathbb{N} \rightarrow \mathbb{R}_{\geq 0}$, which returns the reward to the agent enacting role $i$ in the game once all players perform a joint action $\vec{a}$ at time $t+1$.\footnote{We provide an example of empirical reward function for our traffic scenario in Section \ref{subsec:convergence}.} 

Back with the example of game $G_a$, say that at a given time $t$ the cars have played this game three times so far. In the first two game plays, the cars played with norm combination $\vec{n}=\langle n_4,n_4 \rangle$ and performed a joint action $\vec{a}=\langle stop,stop \rangle$, thus obtaining reward 0.7. In the third game play, the cars played with norm combination $\vec{n}=\langle n_1,n_1 \rangle$ and performed a joint action $\vec{a}=\langle go,go \rangle$, thus getting reward 0.
Hence, the memory of the game at time $t$ is $H=\langle \langle \langle n_4,n_4 \rangle, \langle 0.7, 0.7 \rangle \rangle, \langle \langle n_4,n_4 \rangle, \langle 0.7, 0.7 \rangle \rangle, \langle \langle n_1,n_1 \rangle, \langle 0, 0 \rangle \rangle \rangle$. Say that we want to compute the payoffs of norm combination $\langle n_4,n_4 \rangle$. First, we will retrieve from $H$ the sequence of rewards of $\langle n_4,n_4 \rangle$ up to time $t$, that is, $\langle \langle 0.7, 0.7 \rangle, \langle 0.7, 0.7 \rangle\rangle$. Note that $k=2$, where $k$ is the length of the history of rewards for $\langle n_4,n_4 \rangle$. Let us consider a discount factor $\beta=0.9$, which implies that the $j$-th reward of the sequence is $90\%$ as valuable for the payoff computation as the $(j+1)$-th one. We will compute the payoffs of this norm combination for each role $i \in \{1,2\}$ as:

\begin{flalign*}
\rho_i(\langle n_4,n_4 \rangle, H) & = \dfrac{1}{0.9^1+0.9^0} \cdot \Big[0.7 \cdot 0.9^1 + 0.7 \cdot 0.9^0 \Big] = \mathbf{0.7} 
\end{flalign*}

\subsubsection{Replicating norms}
\label{subsubsec:replicating}

As previously detailed, norm replication is the process of computing the fitness of each norm (Definition \ref{def:fitness} in Section \ref{subsec:problem}), and then making its frequency grow proportionally to its fitness. The \acro{ENSS} computes a norm's fitness similarly to the way a strategy's fitness is computed in EGT (Section \ref{sec:background}). Given a norms game, the fitness of a norm $n$ will depend on: 

\begin{enumerate}
\item the \textit{payoff} that an agent can expect to obtain when using norm $n$ to play the game against other agents with possibly different applicable norms in the game; and 
\item the \textit{probability} that the agent encounters these agents, which can be computed in terms of the frequencies of the norms applicable to these agents in the game. 
\end{enumerate}

Intuitively, if an agent obtains a high payoff once she plays a game against agents with a highly frequent norm, then the agent will be very likely to encounter an agent that uses that norm to play that game, and hence to get a high fitness. Conversely, the same agent will very likely get a low fitness if she is highly likely to interact with agents against whom she always gets a low payoff. 

As an example, let us consider that a car repeatedly plays game $G_a$ in Figure \ref{fig:game1} by using norm $n_1$ in Table \ref{tab:norms}. According to this norm, the car will never give way. This car will yield a high payoff when playing against cars that have applicable norm $n_4$, since this norm obliges a car to always give way. This occurs when the combination of norms used in the game is either $\langle n_1, n_4 \rangle$ (with our car playing role 1), or $\langle n_4,n_1 \rangle$ (with our car playing role 2). Conversely, this car will derive a low payoff when it interacts with cars that have $n_1$ (since both will go forward and collide), namely when the combination of norms used in the game is $\langle n_1,n_1 \rangle$. Now, say that the number of cars with norm $n_4$ doubles the number of cars with norm $n_1$. Then, our car will be twice as likely to play against cars that have $n_4$, and hence to obtain a higher fitness. 

Given a norms game $NG$, we compute the fitness of a norm $n$ at time $t$ as the average payoff $\rho_i$ to an agent once she uses norm $n$ to play $NG$, for each role $i$ and each combination of norms applicable to the players of $NG$. Formally:
\begin{equation}
f(n,NG,t) = \sum\limits_{i=1}^{|R|}  \ \ \sum\limits_{\vec{n} \in N_G^{|R|} \ | \ \vec{n}_i = n}  \rho_i(\vec{n}, H) \cdot p(\vec{n},t) 
\label{eq:fitness}
\end{equation}
where:
\begin{itemize}
\item $N_G^{|R|}$ is the set of all norm combinations that the agents playing the game can employ; 
\item $\vec{n}$ is a norm combination and $\vec{n}_i = n$ is the norm employed by the agent playing role $i$;
\item $\rho_i(\vec{n}, H)$ is the payoff to role $i$ when the agents play with norm combination $\vec{n}$, computed based on the game's memory $H$ up to time $t$; and
\item $p(\vec{n},t)$ is the joint frequency of the norms in $\vec{n}$ in the normative systems of the players.
\end{itemize}

We compute the joint frequency of the norms in $\vec{n}$ in the normative systems of the players of $NG$ at a given time $t$ as:
\begin{equation}
p(\vec{n},t) = \prod\limits_{n \in \vec{n}} F(n,t) = \frac{|\{ag \in Ag \ | \ n \in \Omega_{ag}\}|}{|Ag|}
\end{equation}
where $F(n,t) \in [0,1]$ is the frequency of norm $n$ at time $t$, namely the proportion of agents whose normative systems contain $n$ at time $t$. 

Next, those norms whose fitness is higher than average fitness will become more frequent, while those below average will decrease. This will be captured by a replication equation that we compute as follows. Given a norms game $NG$, we update the frequency of a norm $n \in N_G$ as:

\begin{equation}
F(n,t+1) = F(n,t) + F(n,t) \cdot \big[f(n,NG,t)-\Theta\big]
\label{eq:npfrequpdate}
\end{equation}
\noindent
where $\Theta$ is the average fitness at time $t$ of all the norms applicable in $NG$, computed as:
\begin{equation}
\Theta = \sum_{n \in N_G} f(n,NG,t) \cdot F(n,t) 
\label{eq:avgutility}
\end{equation}

Notice that equations \ref{eq:npfrequpdate} and \ref{eq:avgutility} are the counterparts of \ref{eq:replicationHawk}-\ref{eq:replicationDove} and \ref{eq:avgFitnessEgt} introduced in Section \ref{subsec:replicator} to describe the replicator dynamics of EGT.

As an example, let us compute the fitness of the norms in Table \ref{tab:norms} for game $G_a$. For simplicity, let us consider that only norms $n_1$ and $n_2$ are available to the cars. At the outset, half of the cars have $n_1$ in their normative systems, while the rest of cars have $n_2$. Thus, it follows that the cars can employ the following norm combinations to play the game: $\langle n_1,n_1 \rangle$, $\langle n_1,n_2 \rangle$, $\langle n_2,n_1 \rangle$, and $\langle n_2,n_2 \rangle$. The joint probability of each of these combinations is 0.25 (e.g., $p(\langle n_1,n_1 \rangle,t) = F(n_1,t) \cdot F(n_1,t) = 0.5 \cdot 0.5 = 0.25$). Also, let us consider that at time $t$ our system computes the payoff matrix illustrated in Table \ref{tab:normspayoffs} by means of equation \ref{eq:payoff} (Section \ref{subsec:payoffs}) and based on a memory $H$ of the norms game $N_{G_a}$. Then, we can compute the fitness of norm $n_1$ by using equation \ref{eq:fitness}  as follows:

\begin{flalign*}
f(n_1,G_a,t) &= \rho_1(\langle n_1,n_1 \rangle, H) \cdot p(\langle n_1,n_1 \rangle,t) + \rho_1(\langle n_1,n_2 \rangle, H) \cdot p(\langle n_1,n_2 \rangle,t) + \\ \nonumber
& \ \ \ \ \ \, \rho_2( \langle n_1,n_1 \rangle, H) \cdot p(\langle n_1,n_1 \rangle, t) + \rho_2(\langle n_2,n_1 \rangle, H) \cdot p(\langle n_2,n_1 \rangle,t)\\\nonumber
&= \textbf{0} \cdot 0.25 + \textbf{0} \cdot 0.25 + \textbf{0} \cdot 0.25 + \textbf{1} \cdot 0.25 =\mathbf{0.25} \\\nonumber
\end{flalign*}

Analogously, we compute the fitness of norm $n_2$ as follows:
\begin{flalign*}
f(n_2,G_a,t) &= \rho_1(\langle n_2,n_1 \rangle, H) \cdot p(\langle n_2,n_1 \rangle, t) + \rho_1(\langle n_2,n_2 \rangle, H) \cdot p(\langle n_2,n_2 \rangle, t) + \\ \nonumber
&\ \ \ \ \ \,\rho_2(\langle n_1,n_2 \rangle, H) \cdot p(\langle n_2,n_1 \rangle, t) + \rho_2(\langle n_2,n_2 \rangle, H) \cdot p(\langle n_2,n_2 \rangle,t)
\\ \nonumber
&= \textbf{0.7} \cdot 0.25 + \textbf{0.7} \cdot 0.25 + \textbf{0} \cdot 0.25 + \textbf{1} \cdot 0.25 =\mathbf{0.6} \\
\end{flalign*}

\begin{table}
\centering
\begin{tabular}{cccc}
\multicolumn{1}{l}{} & \multicolumn{1}{l}{} & \multicolumn{2}{c}{\textbf{role 2}} \\
\multicolumn{1}{l}{} & \multicolumn{1}{l}{} & \multicolumn{2}{c}{\textit{\textbf{(right car)}}} \\ \cline{3-4} 
\multirow{3}{*}{\textbf{\begin{tabular}[c]{@{}c@{}}role 1\\ \textit{(left car)}\end{tabular}}} & \multicolumn{1}{c|}{} & \multicolumn{1}{c|}{$n_1$} & \multicolumn{1}{c|}{$n_2$} \\ \cline{2-4} 
 & \multicolumn{1}{|c|}{$n_1$} & \multicolumn{1}{c|}{0,0} & \multicolumn{1}{c|}{0,0} \\ \cline{2-4} 
 & \multicolumn{1}{|c|}{$n_2$} & \multicolumn{1}{c|}{0.7,1} & \multicolumn{1}{c|}{0.7,1} \\ \cline{2-4} 
\end{tabular}
\vspace{0.1in}
\caption{Payoffs computed for norms $n_1$ and $n_2$ after the agents have repeatedly played game $G_a$ in Figure \ref{fig:game1} for a sufficient amount of time. Once the cars use $n_1$ to coordinate, they go forward and collide, getting payoff 0. Only when the car on the left (the one enacting role 1) applies norm $n_2$ (hence stopping) will the cars avoid colliding. In that case, the left car gets payoff 0.7, and the right car gets payoff 1.}
\label{tab:normspayoffs}
\end{table}

Note therefore that norm $n_2$ is more than twice as fit as norm $n_1$. Now, let us replicate both norms. We compute the average fitness of norms $n_1$ and $n_2$ using equation \ref{eq:avgutility} as:
\begin{flalign*}
\Theta &= f(n_1,G_a) \cdot F(n_1) + f(n_2,G_a) \cdot F(n_2) \\
 &= 0.25 \cdot 0.5 + 0.6 \cdot 0.5 = \mathbf{0.425} \\
\end{flalign*}
Since norm $n_2$'s fitness is larger than the average, its frequency in the next generation must increase, while that of $n_1$ must decrease. Specifically:
\begin{flalign*}
F(n_1,t+1) &= F(n_1,t) + F(n_1,t) \cdot \big[f(n_1,G_a,t)-\Theta\big] \\
 &= 0.5 + 0.5 \cdot \big[0.25-0.425\big] = \mathbf{0.4125} \\
\end{flalign*}
\begin{flalign*}
F(n_2,t+1) &= F(n_2,t) + F(n_2,t) \cdot \big[f(n_2,G_a,t)-\Theta\big] \\
 &= 0.5 + 0.5 \cdot \big[0.6-0.425\big] = \mathbf{0.5875} \\
\end{flalign*}
Hence, at time $t+1$, approximately $59\%$ of the agents will adopt $n_2$ in their normative systems, which means that norm $n_2$ will spread. The remaining $41\%$ of the agents will adopt $n_1$, and hence the presence of $n_1$ in the agents' normative systems will shrink.

\subsubsection{Evolutionarily stable normative systems}
\label{subsubsec:stability}

At this point we are ready to provide the stability conditions for norms and normative systems. Analogously to the definition of ESS (see Section \ref{subsec:ess}), we say that a norm is evolutionarily stable if, once a majority of agents use it to coordinate in a game (that is, a majority of agents \textit{have} the norm in their normative systems), there is no alternative norm applicable in the game with which the agents can derive a higher fitness. Consider a MAS such that the majority of its agents have norm $n \in N_G$ in their normative systems to employ when playing a norms game $NG$. Therefore, notice that the frequency of $n$ in the agents' normative systems is close to $1$, namely $F(n,t) \simeq 1$. We say that $n$ is evolutionarily stable when it satisfies one of the following conditions:

\begin{enumerate}

\item \textbf{It is a best response to itself.} Once a whole group of agents play $NG$ by having norm $n$ applicable, no agent can get a higher payoff by using an alternative norm $n'$ (such that $n \neq n'$) to play the game.

\vspace{0.05in}

\item \textbf{It is a best response to any alternative norm.} Considering that condition 1 above does not hold, an agent that plays $NG$ against a group of agents that have any alternative norm $n' \in N_G$ (such that $n \neq n'$) will derive a higher payoff by using norm $n$ to play the game against these agents than by using norm $n'$.
\end{enumerate}

\noindent 
Next, we provide our formal definition of evolutionarily stable norm. 

\begin{mydef}[Evolutionarily stable norm]
\label{def:nstability}
Let $NG=\langle G, N_G, H, P\rangle$ be a norms game with a memory $H$ at a given time $t$, and $n \in N_G$ a norm applicable in $G$. We say that $n$ is evolutionarily stable at time $t$ if, once a majority of agents have $n$ in their normative systems at time $t$, namely once $F(n,t) \simeq 1$, for any alternative norm $n' \in N_G$, $n \neq n'$, one of the two following conditions holds: 
\begin{flalign}
\rho_i(\langle n_1, \ldots, n_m \rangle, H) &> \rho_i(\langle n_1, \ldots, n'_i, \ldots, n_m \rangle, H) \ \mbox{ for all } i \in R \label{eq:esn1}\\
\rho_i(\langle n_1, \ldots, n_m \rangle, H) &= \rho_i(\langle n_1, \ldots, n'_i, \ldots, n_m \rangle, H) \  \mbox{ and } \label{eq:esn2}\\
\rho_i(\langle n'_1, \ldots, n'_m \rangle, H) &< \rho_i(\langle n'_1, \ldots, n_i, \ldots, n'_m \rangle, H) 
\ \mbox{ for all } i \in R \nonumber
\end{flalign}
\end{mydef}
\noindent
where $n_i$ stands for the norm applicable to the agent taking on role $i$, and $n'_i$ is an alternative norm applicable to the agent playing role $i$.

Condition \ref{eq:esn1} means that $n$ is a best response to itself, whereas condition \ref{eq:esn2} means that $n$ is not necessarily a best response to itself, but it is a best response to any alternative norm $n'$. From this definition follows that an evolutionarily stable norm will be a best choice for coordination, since there will be no alternative norm that yields more expected fitness.

Given a set of norms games, we say that a normative system is evolutionarily stable iff all its norms are evolutionarily stable. Formally:

\begin{mydef}[Evolutionarily stable normative system]
Let $\mathcal{NG}$ be a set of norms games and a MAS whose agents are $Ag$. We say that a normative system $\Omega$ is evolutionarily stable at time $t$ iff: (i) for each norms game $NG \in \mathcal{NG}$ there is one and only one norm $n \in \Omega$ and $n \in N_G$ such that $n$ is evolutionary stable from time $t$; and (ii) $\Omega$ is the normative system of all the agents in $Ag$, namely $\Omega_i = \Omega$ for all $ag_i \in Ag$.    
\end{mydef}

Then, if all the agents in a MAS count on the very same normative system, and this is evolutionary stable, no agent will be able to obtain a larger fitness by switching to any alternative normative system. Therefore, complying with its norms will form a rational choice for the agents.

\section{Empirical analysis and results}
\label{sec:empirical}

In this section we empirically evaluate our approach in a simulated traffic scenario. We explore several dimensions. First, we analyse its \textit{convergence}, showing that it manages to converge 100\% of times to ESNSs that avoid collisions as far as cars give a sufficiently high importance to avoiding collisions, namely as far as they are sufficiently \textit{averse to colliding}. We also test the \textit{adaptivity} of our approach, namely its capability to adapt the normative systems it synthesises to the degree of collision aversion of the car population. Finally, we study the \textit{stability} of the normative systems synthesised by our approach upon convergence. We demonstrate that, once all cars abide by an ESNS synthesised by our system, there is no alternative normative system that the agents can be tempted to switch to. 


\subsection{Empirical settings}
\label{subsec:settings}

\begin{figure}
\centering
\includegraphics[width=0.7\linewidth]{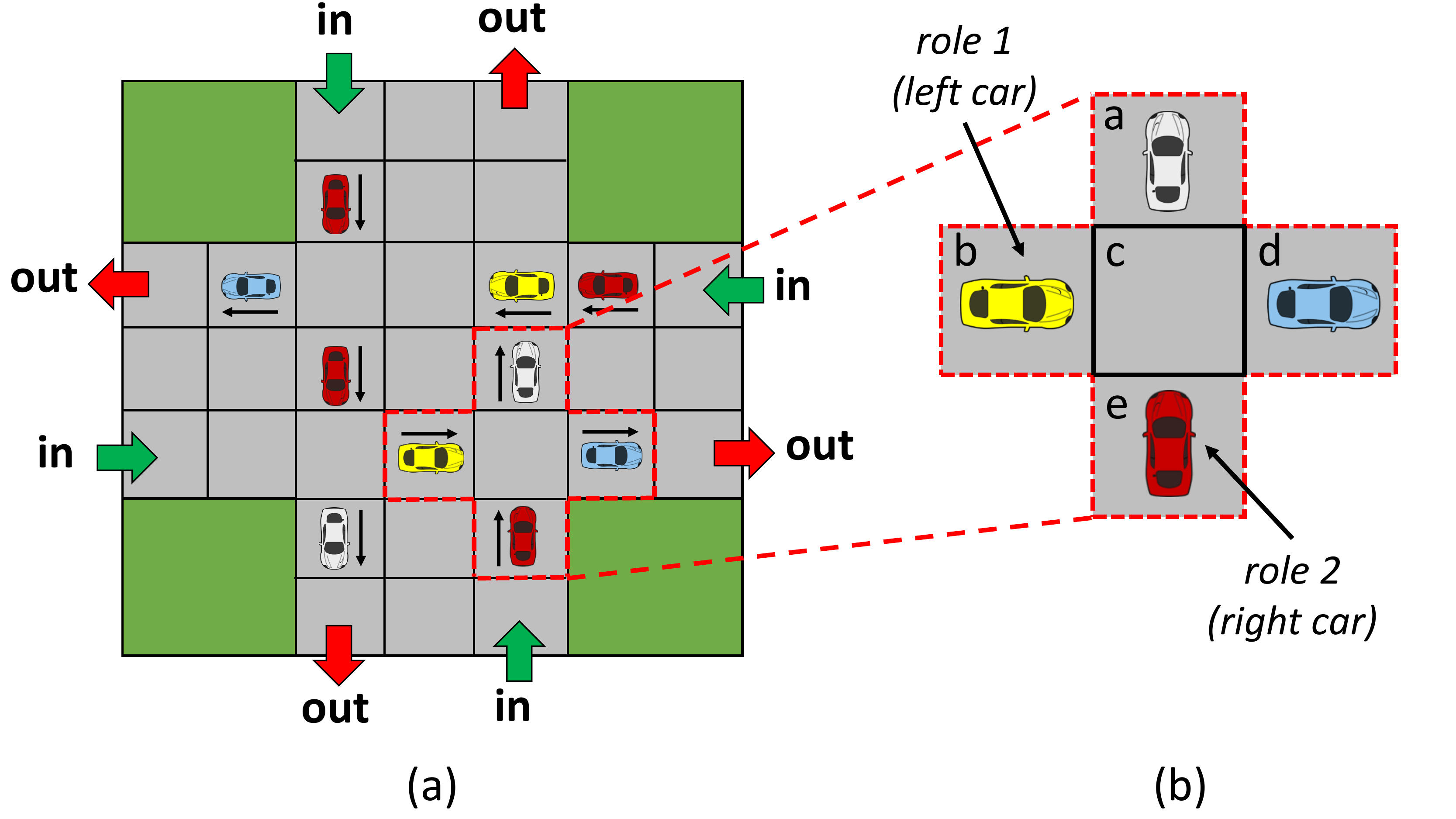}
\caption{(a) Simulated traffic scenario with four entry points (labelled as in), and four exit points (labelled as out); (b) Graphical description of a $2$-role game. Cells \textit{a, b, c} and \textit{d} describe the four positions in front of the car playing role 2 (the red car at cell \textit{e}). Cells \textit{a, c, e} and \textit{d} describe the four positions in front of the car playing role 1 (the yellow car at cell \textit{b}).}
\label{fig:scenario}
\end{figure}

Our experiments consider a discrete simulator of a traffic scenario in which agents are autonomous cars, and the coordination task is to ensure that cars reach their destinations as soon as possible without colliding. Figure \ref{fig:scenario}a illustrates an example of our scenario, composed of two orthogonal roads represented by a $7 \times 7$ grid. At each tick, new cars may enter the scenario from four different entry points (labelled as ``in"), and travel towards one of four exit points (labelled as ``out"). Each car has a limited perception of the scenario and can perceive the four cells in front of it (one cell on its left, two consecutive cells in front, and one cell on its right). Each $m$-role game is described by means of the contents perceived by its players, namely by specifying the four cells in front of each car playing the game. Figure \ref{fig:scenario}b graphically illustrates the description of a $2$-role game played by two of the cars in Figure \ref{fig:scenario}a. Specifically, cells \textit{a, b, c} and \textit{d} describe the four positions in front of the car playing role 2 (the one in cell \textit{e}), and cells \textit{a, c, e} and \textit{d} describe the four positions in front of the car playing role 1 (the one in cell \textit{b}). 

Each experiment consists of a set of simulations that start with a population of agents that have no norms to coordinate (that is, each agent $ag_j \in Ag$ has an empty normative system $\Omega_j=\emptyset$). Simulations run in rounds of 200 ticks. In each round, cars interact in the junction and collisions occur as the simulation goes on\footnote{Whenever two or more cars collide, they remain in the scenario for 5 ticks until they are removed. With this aim we aim to simulate the time that the emergency services require to move away collided cars.}. Our \acro{ENSS} continuously monitors the system, and captures these coordination situations as $m$-role games, creating norms that the cars incorporate to their normative systems (see Section \ref{subsubsec:games}). Over time, the \acro{ENSS} computes the payoffs and the fitnesses of each norm, and evolves norms based on their fitness. We consider that the system has converged whenever the frequency of each norm has not changed during the last 30 rounds ($I=30$). Upon convergence, we consider that a norm $n$ is evolutionarily stable if all the agents have adopted it (that is, if $F(n,t) = 1$). Moreover, if all the cars have adopted the same set of norms and these are evolutionarily stable, then we say that they have converged to an evolutionarily stable normative system. 

\subsection{Convergence analysis}
\label{subsec:convergence}

We first analyse the capability of our approach to synthesise an ESNS that successfully coordinates the cars in avoiding collisions as far as they are sufficiently willing to avoid collisions (namely, as far as they are sufficiently averse to colliding). 
With this aim, we run 1,000 simulations that consider the following empirical reward function (see Section \ref{subsec:payoffs}):
\begin{equation}
\tilde{r}_i(\vec{a},t) = \threepartdef
{0\ \ \ }      	{\mbox{player } i \mbox{ collides at time }t}
{0.7\ \ \ }     {\mbox{player } i \mbox{ avoids collisions but cannot move forward at time }t}
{1\ \ \ } 		{\mbox{player } i \mbox{ avoids collisions and can move forward at time }t} \nonumber
\end{equation}
where $\vec{a}$ is a joint action performed by the players of an $m$-role game $G$ at a given time $t$. 

Thus, a car gets the worst possible reward (reward 0) once it plays a game at time $t-1$ and collides at time $t$. The best possible reward (reward 1) it gets once avoids collisions and can go forward (hence not delaying). Finally, a car gets a less positive reward (reward 0.7) when it has to stop in order to not collide (which, on the other hand, is detrimental to the goal of reaching its destination as soon as possible)\footnote{Note that with this reward function the system will learn payoff matrices that are similar to the one considered in our example game of Table \ref{tab:game} in Section \ref{sec:model}.}. Note that the two rewards for not colliding are significantly higher than the reward for colliding. Thus, cars will give a higher importance to avoiding collisions at the expense of travelling time. In other words, we say that the cars will be highly averse to colliding. 

Finally, we consider a discount factor of 0.8 ($\beta=0.8$) to compute empirical payoffs in Equation \ref{eq:payoff} (Section \ref{subsec:payoffs}).\footnote{We also performed simulations with discount factors that ranged from 1 to 0.1, obtaining similar results.}

\begin{figure}
\centering
\includegraphics[width=1\linewidth]{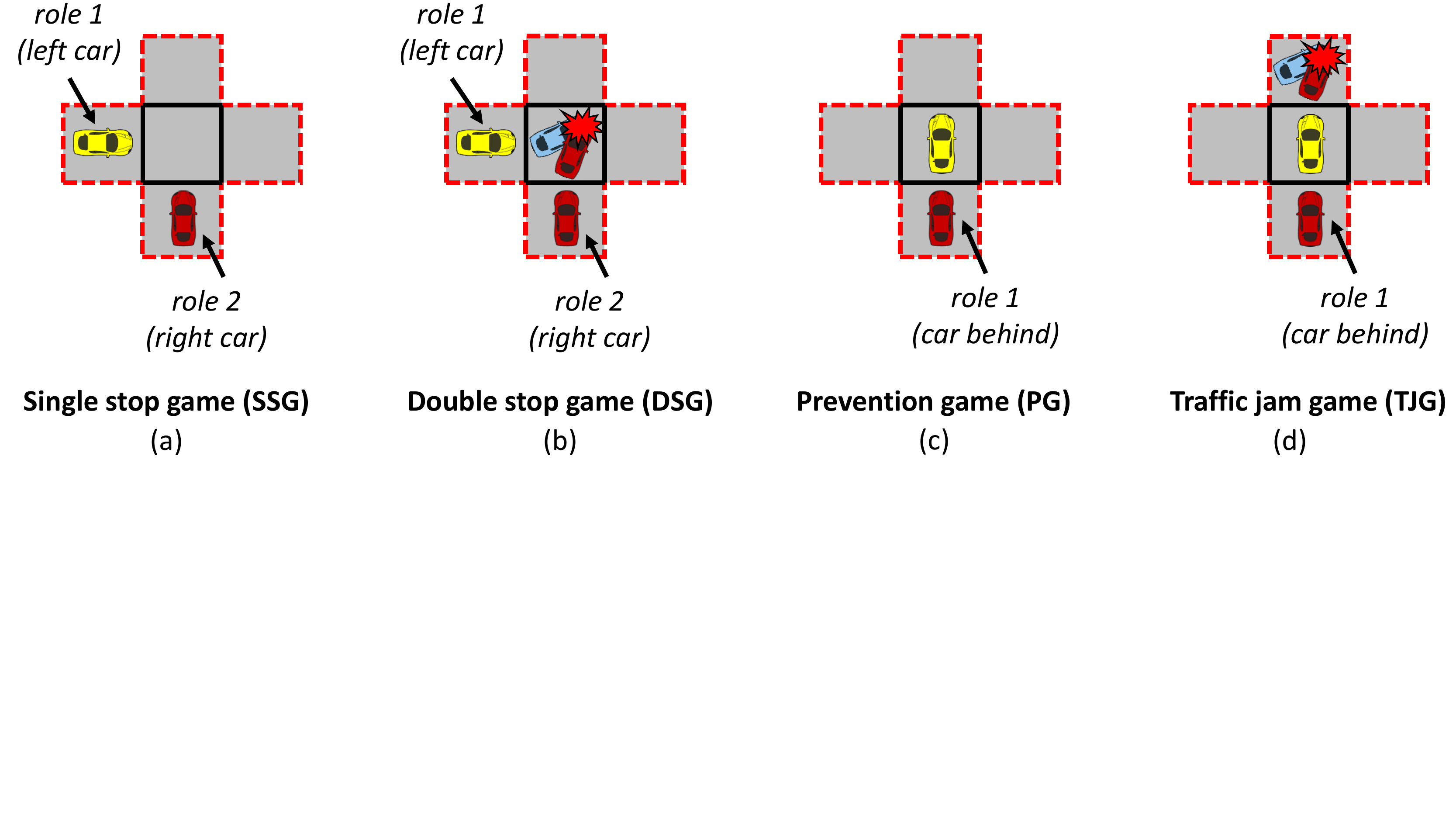}
\vspace{-1.3in}
\caption{Different game types detected in our simulations: (a) a $2$-role \textit{single-stop game} (SSG), in which the best strategy is that one of the cars stops, giving way to the other; (b) a $2$-role \textit{double-stop game} (DSG), in which both cars have to stop in order to not collide; (c) a $1$-role \textit{prevention game} (PG), in which a car (the one at cell \textit{e}) can stop for one tick, thus keeping a security distance with the car in front (the one at cell \textit{c}), or to go forward, assuming the risk of colliding in case the car in front stops; (d) a $1$-role \textit{traffic-jam game} (TJG), in which a car (the one at cell \textit{e}) has to stop in a traffic jam in order to not collide with the car in front.}
\label{fig:gameTypes}
\end{figure}

Out of 1,000 simulations, the \acro{ENSS} takes an average of 54 rounds to converge. During that time, it generates 64 different games that can be grouped into the four categories illustrated in Figure \ref{fig:gameTypes}. The first category (Figure \ref{fig:gameTypes}a), which we call \textit{single-stop games} (SSG), stands for $2$-role games in which the best strategy to avoid collisions and to delay as little as possible is that one of the cars stops, giving way to the other.
Two examples of SSG are the one illustrated in Figure \ref{fig:gameTypes}a, and the one depicted in Figure \ref{fig:scenario}b, which is very similar to the former one but also considers a third and a fourth car in cells \textit{a} and \textit{d}. In general, any variation in cells \textit{a, c} and \textit{d} of a $2$-role game is considered as a different game\footnote{This is due to the assumed non-determinism of the MAS, whereby our system cannot assume that two similar situations will stand for similar games.}. 

The second category, which we call \textit{double-stop games} (DSG), stands for $2$-role games in which both cars need to stop in order to avoid collisions (at the expense of extra travel time). Figure \ref{fig:gameTypes}b shows an example of DSG, in which two cars are waiting for a collision to be moved away. The third category (Figure \ref{fig:gameTypes}c), called \textit{prevention games} (PG), stands for $1$-role games in which a car can go forward, assuming a collision risk in case the car in front stops, or to stop for one time step in order to keep a security distance. The fourth category (Figure \ref{fig:gameTypes}d) we call \textit{traffic-jam games} (TJG). They are similar to prevention games, but in this case a car is stuck in a jam and it can decide whether to stop, doing the queue, or to go forward, colliding with the car in front.

Overall, the \acro{ENSS} detects 12 different single-stop games, 7 double-stop games, 40 prevention games, and 5 different traffic jam games. Table \ref{tab:normsGames} illustrates the six possible types of strategies to coordinate cars in each type of game, along with the average number of simulations that converged to norms that prescribed these strategies. The first four strategies are aimed to regulate $2$-role games (SSGs and DSGs)\footnote{For instance, norms $n_1, n_2, n_3, n_4$ in Table \ref{tab:norms} (Section \ref{subsec:model}) prescribe these strategies to regulate $G_a$.}. They state that a car has to: ``never give way" when playing either role 1 or 2 (no prohibitions imposed); ``give way to the right", i.e., it is prohibited to go when playing role 1 ($\{\langle 1,go \rangle \}$); ``give way to the left" ($\{\langle 2,go \rangle \}$); and ``give way always" ($\{\langle 1,go \rangle ,\langle 2,go \rangle \}$). The last two strategies regulate $1$-role games (PGs and TJGs). Strategy ``go" ($\{\}$) says that a car is free to go forward, and strategy ``stop" ($\{\langle 1,go \rangle \}$) says that a car is prohibited to go (either to keep a security distance with the car in front, or to do a queue). As a matter of fact, strategy ``stop" is the only one that can avoid 100\% of collisions in both PGs and TJGs.

\begin{table}[]
\footnotesize
\centering
\begin{tabular}{l|c|c|c|c|c|c|}
 \multicolumn{7}{c}{\textbf{Norms}}  \\
\cline{2-7}
 & \textit{\textbf{\begin{tabular}[c]{@{}c@{}}``never give way"\\ $\{ \}$ \end{tabular}}} & \textit{\textbf{\begin{tabular}[c]{@{}c@{}}``give way\\to the right"\\ $\{\langle 1,go \rangle \}$ \end{tabular}}} & \textit{\textbf{\begin{tabular}[c]{@{}c@{}}``give way\\to the left"\\ $\{\langle 2,go \rangle \}$\end{tabular}}} & \textit{\textbf{\begin{tabular}[c]{@{}c@{}}``give way always"\\ $\{\langle 1,go \rangle ,\langle 2,go \rangle \}$\end{tabular}}} & \textit{\textbf{\begin{tabular}[c]{@{}c@{}}``go"\\ $\{ \}$\end{tabular}}} & \textit{\textbf{\begin{tabular}[c]{@{}c@{}}``stop"\\ $\{\langle 1,go \rangle \}$ \end{tabular}}} \\ \hline
\multicolumn{1}{|l|}{\textbf{SSG}} & 0\% & \textbf{49\%} & \textbf{51\%} & 0\% & --- & --- \\ \hline
\multicolumn{1}{|l|}{\textbf{DSG}} & 0\% & 0\% & 0\% & \textbf{100\%} & --- & --- \\ \hline
\multicolumn{1}{|l|}{\textbf{PG}} & --- & --- & --- & --- & \textbf{90\%} & 10\% \\ \hline
\multicolumn{1}{|l|}{\textbf{TJG}} & --- & --- & --- & --- & 0\% & \textbf{100\%} \\ \hline
\end{tabular}
\vspace{0.2in}
\caption{Different types of norms to coordinate the cars in each type of game, along with the average number of times that cars converged to adopting each type of norm. The first four norms regulate SSGs and DSGs (2-role games). The last two norms regulate PGs and TJGs (1-role games).}
\label{tab:normsGames}
\end{table}

On average, cars converge to adopting an evolutionarily stable norm that avoids collisions in 100\% of SSGs and DSGs. Specifically, in SSGs cars adopt 49\% of times a norm to ``give way to the right", and the remaining 51\% of times they adopt a ``give way to the left" norm. As for DSGs, cars adopt a norm prescribing a ``give way always" strategy 100\% of times. In PGs, cars adopt 90\% of times a norm prescribing a ``stop" strategy in order to keep a security distance, and the remaining 10\% of times they adopt the ``go" norm (hence assuming the risk of colliding). This happens because in PGs, the car in front (e.g., the car in cell \textit{e} in Figure \ref{fig:gameTypes}c) does not always stop, and hence the collision risk of proceeding is lower than 100\%. Since the reward for going forward and not colliding (reward 1) is higher than that of stopping and not colliding (reward 0.7), sometimes cars prefer to assume the collision risk and adopting a ``go" norm in order to save travel time. Conversely, in TJGs cars adopt a norm prescribing ``stop" 100\% of times, since in those games the risk of colliding once going forward is 100\%. 

\begin{figure}
\centering
\includegraphics[width=0.9\linewidth]{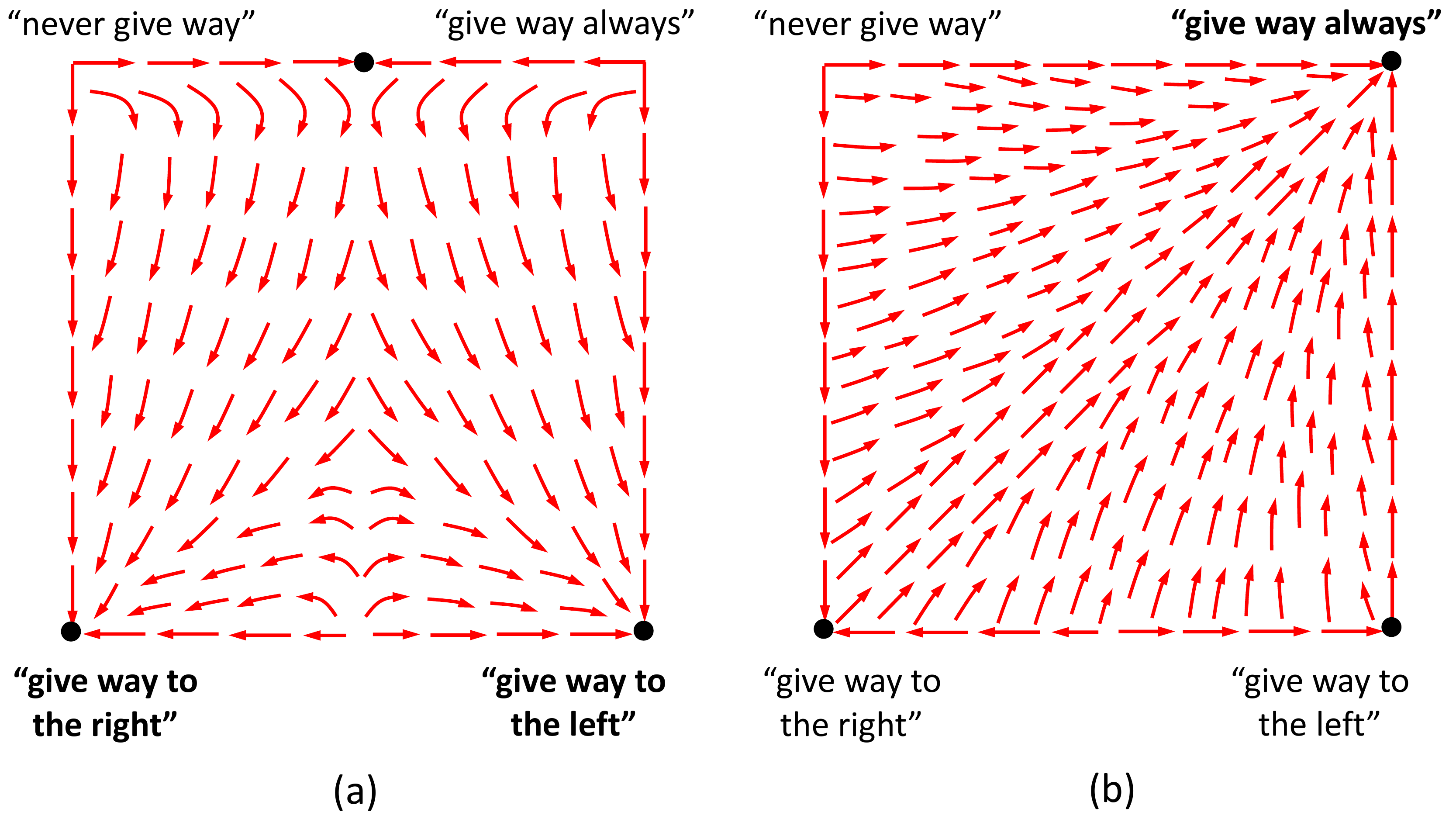}
\caption{Evolutionary dynamics of norm adoption in SSGs (a) and DSGs (b). Each square represents the possible frequency distributions of the first four norms in Table \ref{tab:normsGames}, which establish that a car has to: ``never give way", ``give way to the right", ``give way to the left", and ``give way always". Arrows represent the gradient of norm adoption for each norm distribution, i.e., the most likely trajectory in terms of norm adoption that a population with a given norm distribution will follow. Figure (a) shows the dynamics of SSGs, in which any norms prescribing strategies ``give way to the right" or ``give way to the left" are evolutionarily stable. Also, there is an evolutionarily stable state with a polymorphic population whose 50\% of cars ``never give way", and the remaining 50\% ``give way always". This state can be reached when cars do not have available norms to give way to either the left or the right. Figure (b) shows the dynamics of DSGs, in which the evolutionarily stable norms are those prescribing strategy ``give way always". Also, there are two evolutionarily stable states in which cars only have available either norms to ``give way to the left" or norms to ``give way to the right".}
\label{fig:dynamics}
\end{figure}

Figure \ref{fig:dynamics} shows the evolutionary dynamics of norm adoption for SSGs (Figure \ref{fig:dynamics}a) and DSGs (Figure \ref{fig:dynamics}b). Each square represents the possible frequency distributions of norms prescribing strategies to ``never give way", ``give way to the right", ``give way to the left" and ``give way always". For instance, the top-left corner represents a population in which 100\% of cars adopt a ``never give way" norm, and the middle point of the square represents a population in which the four norms are 25\% frequent. Arrows represent the gradient of norm adoption for each norm distribution, i.e., the most likely trajectory in terms of norm adoption that a population with a given norm distribution will follow.

In SSGs, norms to ``give way to the right" or to ``give way to the left" are the only evolutionarily stable ones. Both norms are attractor points of the norm evolution process (indicated with big black dots). If the mass of cars giving way to the right is bigger than the mass of cars giving way to the left, then the whole population will tend to give way to the right in order to synchronise. There is also an evolutionarily stable state in which 50\% of cars adopt a ``never give way" norm, which the remaining 50\% of cars compensate adopting a ``give way always" norm. This state can be reached whenever cars do not have available norms to give way to a side (which could invade the population because they are fitter). As for DSGs, norms to ``give way always" are the only evolutionarily stable ones. Thus, no matter what the initial norm distribution is, as long as at least one car adopts a ``give way always" norm, its fitness will be higher than that of any other car, and the whole population will eventually adopt its norm.

\subsection{Adaptivity analysis}
\label{subsec:adaptivity}

Next, we analyse the capability of our \acro{ENSS} to adapt norm synthesis to the characteristics of the population to regulate. With this aim, we run simulations with populations that have different degrees of aversion to colliding. We model these populations by considering a collection of empirical reward functions depicted in Table \ref{tab:rewards}. Each function returns 0 once a car collides at time $t$, and 1 once a car goes forward without colliding at time $t$. These functions differ in the reward given to the cars once they \textit{stop} to avoid collisions, which balances their ``hurry" to get to their destinations with their willingness to avoid collisions. If this reward is low (e.g., $\tilde{r}^0_i$), then the cars prefer not to stop in order to not being delayed, even if it implies a collision risk. Then, we say that cars will have a low collision aversion degree. As this reward increases, cars will have a lower aversion to stopping, which can be interpreted as a higher aversion to colliding. 

\begin{table}[]
\centering
\begin{tabular}{lccccccccccc}
 \textbf{Outcome} & \multicolumn{1}{c}{$\tilde{r}^0_i$} & \multicolumn{1}{c}{$\tilde{r}^1_i$} & \multicolumn{1}{c}{$\tilde{r}^2_i$} & \multicolumn{1}{c}{$\tilde{r}^3_i$} & \multicolumn{1}{c}{$\tilde{r}^4_i$} & \multicolumn{1}{c}{$\tilde{r}^5_i$} & \multicolumn{1}{c}{$\tilde{r}^6_i$} & \multicolumn{1}{c}{$\tilde{r}^7_i$} & \multicolumn{1}{c}{$\tilde{r}^8_i$} & \multicolumn{1}{c}{$\tilde{r}^9_i$} & \multicolumn{1}{c}{$\tilde{r}^{10}_i$} \vspace{0.03in} \\ \hline
\multicolumn{1}{|l|}{\textit{Collides}} & \multicolumn{1}{c|}{0} & \multicolumn{1}{c|}{0} & \multicolumn{1}{c|}{0} & \multicolumn{1}{c|}{0} & \multicolumn{1}{c|}{0} & \multicolumn{1}{c|}{0} & \multicolumn{1}{c|}{0} & \multicolumn{1}{c|}{0} & \multicolumn{1}{c|}{0} & \multicolumn{1}{c|}{0} & \multicolumn{1}{c|}{0} \\ \hline
\multicolumn{1}{|l|}{\textit{Stops \& avoids collisions}} & \multicolumn{1}{c|}{0} & \multicolumn{1}{c|}{0.1} & \multicolumn{1}{c|}{0.2} & \multicolumn{1}{c|}{0.3} & \multicolumn{1}{c|}{0.4} & \multicolumn{1}{c|}{0.5} & \multicolumn{1}{c|}{0.6} & \multicolumn{1}{c|}{0.7} & \multicolumn{1}{c|}{0.8} & \multicolumn{1}{c|}{0.9} & \multicolumn{1}{c|}{1} \\ \hline
\multicolumn{1}{|l|}{\textit{Goes \& avoids collisions}} & \multicolumn{1}{c|}{1} & \multicolumn{1}{c|}{1} & \multicolumn{1}{c|}{1} & \multicolumn{1}{c|}{1} & \multicolumn{1}{c|}{1} & \multicolumn{1}{c|}{1} & \multicolumn{1}{c|}{1} & \multicolumn{1}{c|}{1} & \multicolumn{1}{c|}{1} & \multicolumn{1}{c|}{1} & \multicolumn{1}{c|}{1} \\ \hline
\end{tabular}
\vspace{0.2in}
\caption{Empirical reward functions to model populations with different degrees of collision aversion. The lower rewards (e.g., $\tilde{r}^0_i, \tilde{r}^1_i$) represent populations with lower aversion to colliding. The higher rewards (e.g., $\tilde{r}^9_i, \tilde{r}^{10}_i$) represent populations with higher aversion to colliding.}
\label{tab:rewards}
\end{table}

\begin{figure}
\centering
\includegraphics[width=0.85\linewidth]{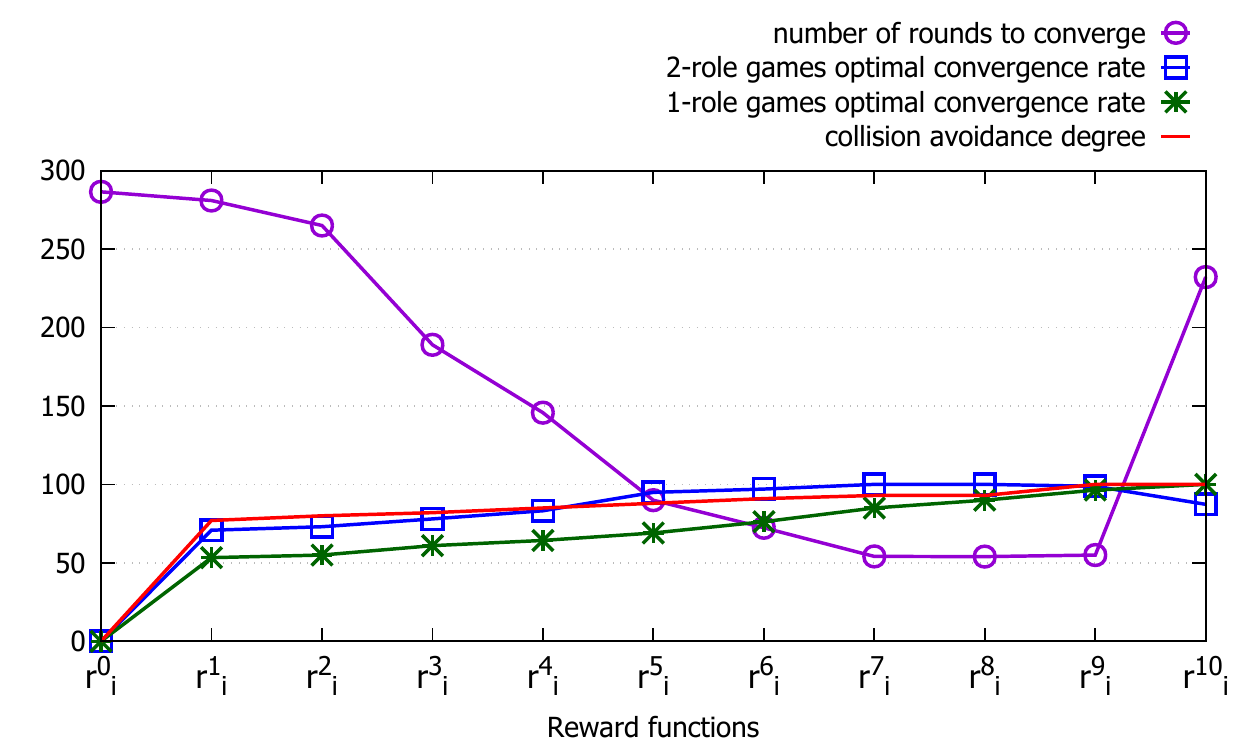}
\caption{Averaged results of 1,000 simulations with a 0.8 discount factor ($\beta=0.8$) and different degrees of collision aversion. The $x$-axis represents different reward functions, which allow to model the different collision aversion degrees. For instance, $\tilde{r}^0_i$ represents a population with null collision aversion, and $\tilde{r}^{10}_i$ represents a population with total collision aversion. The $y$-axis shows: (1) the number of rounds the simulations required to converge; (2) the frequency with which cars converge optimally in $2$-role games; (3) the frequency with which cars converge optimally in $1$-role games; and (4) the collision avoidance rate during the last round of the simulation (once agents had converged to an ESNS).}
\label{fig:adaptivity}
\end{figure}

We run 1,000 simulations for each reward function with a 0.8 discount factor ($\beta=0.8$). Figure \ref{fig:adaptivity} shows averaged results of all simulations. The $x$-axis depicts the different empirical reward functions (collision aversion degrees), and the $y$-axis shows:
\begin{itemize}
\item the number of rounds that the \acro{ENSS} requires to converge.
\item the average frequency with which cars optimally converge in $2$-role games. That is, the frequency with which they converge to a norm like ``give way to the right" or a norm like ``give way to the left" in SSGs, and to a norm like ``give way always" in DSGs.
\item the average frequency with which cars optimally converge $1$-role games. That is, the frequency with which they converge to ``stop"  in PGs and TJGs (which are the only strategies that allow to avoid 100\% of collisions). 
\item the collision avoidance rate during the last round of the simulation (once the simulation has converged and the cars have adopted an ESNS). 
\end{itemize}
With null collision aversion ($\tilde{r}^0_i$), simulations take a high number of rounds to converge (286 rounds). This happens because the rewards for colliding and for stopping are equal, and hence the fitness of the norms that cause collisions (i.e., those prescribing ``never give way") and those that prohibit to go in order to avoid collisions (e.g., ``give way to the right") are similar. Consequently, cars take a long time to decide which norm to adopt. Upon convergence, cars adopt ESNSs containing ``never give way" norms, which avoid 0\% of collisions. 

As the collision aversion increases, simulations take less number of rounds to converge, and cars adopt norms that prohibit to go more frequently. For low collision aversions ($\tilde{r}^1_i$ to $\tilde{r}^3_i$), simulations still take a high number of rounds (281, 265 and 189 rounds, respectively), but cars adopt ESNSs that avoid up to 80\% of collisions. Specifically, cars converge optimally up to 73\% of times in $2$-role games, and up to 55\% of times in $1$-role games. The reason that this frequency is slightly lower in $1$-role games is because in prevention games (PG) cars not always collide once they choose to go forward. Hence, cars occasionally converge to norm ``go", which cannot fully avoid collisions. For middle and high collision aversion degrees ($\tilde{r}^4_i$ to $\tilde{r}^9_i$), the number of rounds necessary to converge decreases significantly. The best results are given by functions $\tilde{r}^7_i$ and $\tilde{r}^8_i$, with which cars optimally converge in the 100\% of $2$-role games \--- but still, they optimally converge up to 90\% of $1$-role games, hence avoiding up to 93\% of collisions. 

With total collision aversion ($\tilde{r}^{10}_i$), the number of rounds necessary to converge increases again. This happens because the reward for stopping and not colliding and the reward for going forward and not colliding are equal. Hence, the fitness of all the norms that avoid collisions \--- either by prohibiting one role to go, or by prohibiting both roles to go \--- are similar. In consequence, cars need extra time to decide which norm to adopt. Upon convergence, cars adopt ESNSs containing only ``give way always" norms to coordinate in SSGs and DSGs (hence converging optimally for DSGs, but not for SSGs). As a result, cars remain stopped indefinitely and 100\% of collisions are avoided. It turns out that cars are so afraid of colliding that they do not mind to stay still indefinitely in order to avoid collisions.

\subsection{Stability analysis}
\label{subsec:stability}

Finally, we analyse the stability of the normative systems synthesised by our approach upon convergence. With this aim, we perform 100 simulations that start with a population in which 100\% of agents abide by an ESNS of those synthesised in the experiment of Section \ref{subsec:convergence}, which we will call $\Omega^*$. Each simulation lasts 400 rounds, and each round lasts 200 ticks. In each round, the system creates random agents that abide by a normative system with alternative norms for each possible game. Then, we let agents interact. At the end of each round, the system performs norm replication. As considered in the literature in EGT \cite{smith1988evolution}, we consider that our normative system is an ESNS if none of its norms can be invaded by any alternative norm. That is, if in every single round, the fitness of the mutant norms is lower than the fitness of the norms in $\Omega^*$ (and hence, the invader norms cannot grow in frequency). Thus, if our normative system is an ESNS, the agents will end up adopting it at the end of every single simulation. 

\begin{figure}
\centering
\includegraphics[width=0.76\linewidth]{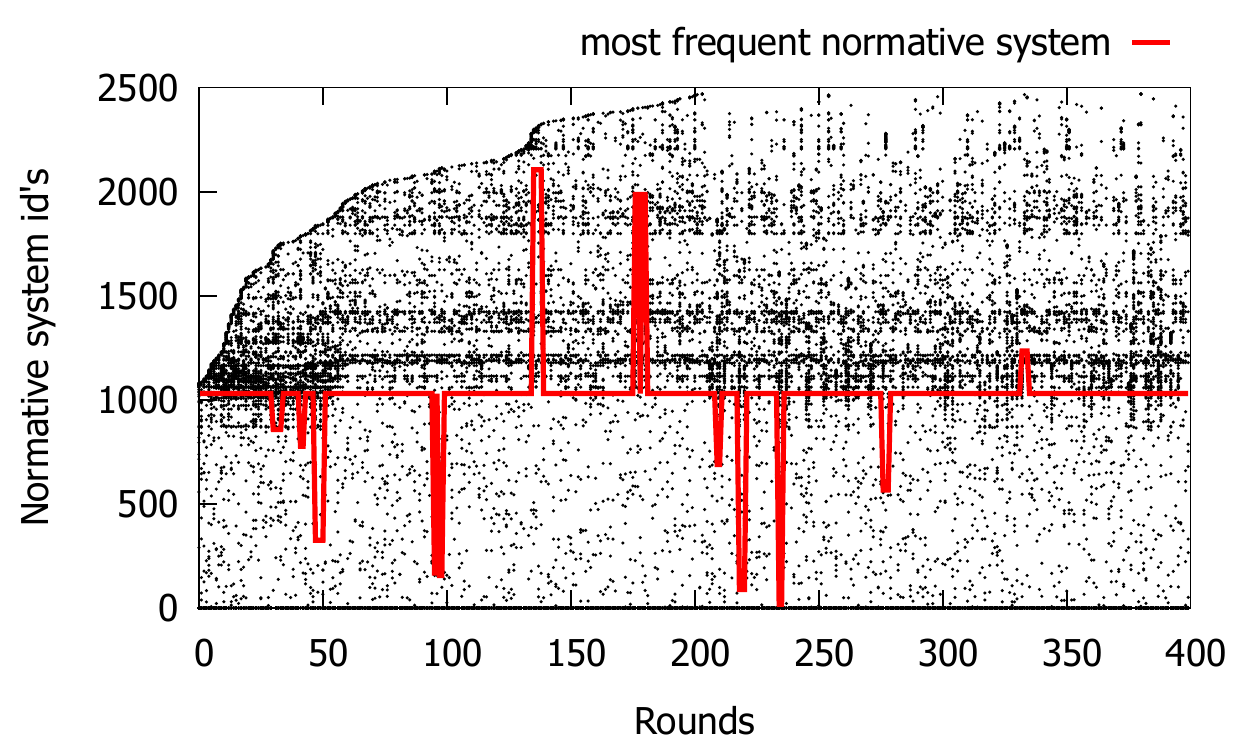}
\vspace{-0.05in}
\caption{Competition of an ESNS $\Omega^*$ (represented with id 1000) against mutant normative systems in one simulation. The $x$-axis illustrates the different rounds of the simulation. The $y$-axis illustrates the id's of the different normative systems generated during the simulation. Black dots represent the creation of mutant normative systems with a certain id at a given round. The red line illustrates the id of the most frequent normative system.}
\label{fig:stability}
\end{figure}

In 100\% of simulations, the cars ultimately adopted normative system $\Omega^*$. Figure \ref{fig:stability} illustrates the dynamics of one of these simulations. The $x$-axis shows the different rounds of the simulation, and the $y$-axis depicts the id's of the normative systems created over time. Black dots represent mutant normative systems created in each round (with which $\Omega^*$ has to compete), and the red line indicates the id of the most frequent normative system. For the sake of clarity, we represent $\Omega^*$ as the normative system with id 1,000. After 200 rounds, the simulation created 2,500 different mutant normative systems. Upon round 400, normative system $\Omega^*$ remained stable most of the time. In punctual rounds, the simulation generated a high number of mutant normative systems, making the frequency of $\Omega^*$ to go below stability. But, after a few rounds, $\Omega^*$ replicated and became again the most frequent normative system. Upon round 400, the cars converged to adopting $\Omega^*$, thus demonstrating that $\Omega^*$ is a best choice for the agents.

\section{Related work}
\label{sec:rw}

Broadly speaking, research on norm synthesis can be classified into two main strands of work: \textit{off-line design}, and \textit{on-line synthesis}. Pioneered by Shoham and Tennenholtz \cite{ShohamT95}, off-line design aims at designing the norms that will coordinate the agents before they start to do their work \cite{ShohamT95,fitoussi00choosing,van2007social,corapi2011normative}. This approach is arguably less flexible, since norms are typically hard-wired into the agents' behaviours and cannot be adapted over time.

Alternatively, on-line synthesis studies how norms can come to exist while the agents interact at runtime. Most on-line approaches focus on investigating how norms \textit{emerge} from agent societies through an iterated process whereby agents tend to adopt best-performing strategies \cite{axelrod1986evolutionary,ShohamT92,walker1995understanding,delgado2002emergence,villatoro2011self,airiau2014emergence}. Alternatively, recent work by Morales et. al. approached on-line synthesis by employing designated agents that observe agents' interactions at runtime and generate norms aimed at resolving conflict situations \cite{morales2013automated,morales2014minimality,morales2015liberality}. Later on, Mashayekhi et. al. \cite{mashayekhi2016silk} extended this work by proposing a hybrid mechanism in which norms are synthesised by designated agents, and the agent society iteratively selects and spread best-performing norms, ultimately adopting most successful ones. 

The closest approach to our work in the literature is that of norm emergence, and particularly the study of norm evolution and stability. One of the pioneer works on this approach is the one by Axelrod \cite{axelrod1986evolutionary,axelrod1997complexity}, which studies how norms evolve and emerge as stable patterns of behaviour of agent societies. Axelrod considers a game-theoretic setting in which agents repeatedly play a single two-player game by employing different strategies. The strategies that allow the agents to achieve better personal results prosper and spread. A (stable) norm is said to have emerged once a majority of agents abide by the same strategy that is sustained over time. 

Subsequently, many researchers have studied norm emergence by employing the game theoretic approach. In \cite{sethi1996evolutionary}, Rajiv Sethi extended the work of Axelrod and studied how social norms of vengeance and cooperation emerge within agent societies. With this aim, Sethi incorporated the solution concept of evolutionarily stable strategy (ESS) and the principle of replicator dynamics from evolutionary game theory (EGT) \cite{smith1988evolution}. Again, this work considers that the agents play a single two-player game. Shoham and Tennenholtz \cite{shoham1997emergence} introduced a framework for the emergence of social conventions as points of (Nash) equilibria in \textit{stochastic} settings. They introduced a natural strategy-selection rule whereby the agents eventually converge to rationally acceptable social conventions. 

Later, Sen and Airiau proposed in \cite{sen2007emergence} a \textit{social learning} model whereby agents can learn their policies and norms can emerge over repeated interactions with multiple agents in two-player games. Many works have considered this model to study further criteria that affect to the emergence of norms. Of these, the closest to our work is perhaps the one by Sugawara et. al. \cite{sugawara2011emergence,sugawara2014emergence}, in which conflict situations are characterised as  Markov games, and a model is introduced to evolve norms that successfully coordinate the agents in these games. 

More recent work (such as that by Santos et. al. \cite{santos2016evolution}) studies how cooperation norms can emerge once the agents can explore alternative strategies, i.e. they have arbitrary exploration rates. They show that cooperation emergence depends on both the exploration rate of the agents and the underlying norms at work. Similarly, Soham et. al. \cite{de2017understanding} introduce an EGT-based model to study how norms change in agent societies in terms of the \textit{need for coordination} and the agents' exploration rate. They show that societies with high needs for coordination tend to lower exploration rates and higher norm compliance, while societies with lower coordination needs lead to higher exploration rates. Also, Lorini et. al. \cite{lorini2015long} introduce a model for the evolution and emergence of fairness norms in relation to the degree of sensitivity (internalisation) of the agents to these norms. They show that, in the long term, the higher the sensitivity of the agents to norms, the more beneficial for the social welfare.

From Axelrod \cite{axelrod1986evolutionary,axelrod1997complexity} to Lorini \cite{lorini2015long}, our approach is different to all the aforementioned works for several reasons. First, most previous works consider that the agents play a \textit{single} game whose payoffs are known beforehand. Unlike them, our work considers a \textit{non-deterministic} setting in which the games (i.e., the conflict situations) that the agents might engage in are initially unknown, likewise the payoffs of these games. We provide a framework to perform runtime detection of the multiple games that the agents might play, and to learn their payoffs based on the rewards to the agents once they repeatedly play each game over time. Our framework allows to evolve norms based on their capability to coordinate agents in different games, and leads agents to converge to normative systems that maximise their fitness. Moreover, to the best of our knowledge, our framework is the first one in introducing the analytical concept of evolutionarily stable normative system (ESNS), which allows to assess the combinations of norms that the agents are likely to adopt. 

\section{Conclusions}
\label{sec:conclusions}

In this work we introduced a framework for the synthesis of evolutionarily stable normative systems (ESNS) for non-deterministic settings. Our framework allows to synthesise sets of norms that populations of rational agents are likely to adhere by. With this aim, our framework carries out an evolutionary game theoretic process inspired in evolutionary game theory (EGT), whereby the agents tend to adopt the norms that are fittest to coordinate them in strategic situations. Our framework assumes no previous knowledge about the coordination situations that the agents can engage in, neither about their potential outcomes. Instead, it learns these by letting the agents interact, detecting the situations in which they need coordination, and modelling them as games of which it iteratively learns their payoffs. Norms that are more useful to coordinate the agents in these games prosper and spread, and are ultimately adopted by the agents. The outputs of this evolutionary process are sets of norms that, once adopted by an entire agent population, no agent can benefit from switching to alternative norms (and hence, we say that they are evolutionarily stable). 

We provided evidence of the quality and the relevance of our approach through an empirical evaluation in a simulated traffic scenario. We showed that our framework allows cars to converge 100\% of times to normative systems that successfully avoid car collisions as far as they are sufficiently willing to avoid collisions. We illustrated the capability of our approach to adapt norm synthesis to the preferences of the agent population, showing that the agents will tend to adopt different types of normative systems as their preferences change. In this sense, we believe that our framework provides a valuable decision support tool for policy makers. It can be used to make predictions about the type of normative systems that a given population will end up adopting, and to assess whether a particular population will accept a given normative system. 

Note that our model for norm evolution is described in terms of the \textit{individual} goals of the agents. Thus, the norms that will be more likely to spread are those that are useful to the agents from their point of view. As previously mentioned, this can be very useful in order to predict which norms a given population will ultimately adopt. Nevertheless, our framework cannot be employed to synthesise norms that achieve a \textit{global} goal that is not aligned with the individual goals of the agents. For instance, cars will only adopt norms that avoid collisions as far as they do not want to collide. Therefore, the sensible next step is to extend our model in order to be able to synthesise evolutionarily stable norms that accomplish a coordination task even when coordination is not in the interests of the agents. For instance, to synthesise norms that avoid collisions even when cars do not mind to collide. With this aim, we plan to include a deterrence mechanism in our model by incorporating sanctions in norms so that coordination can emerge even when the agents' goals are not aligned with the global system goals.

\def\UrlBreaks{\do\/\do-}

\bibliographystyle{abbrv}
\bibliography{NormSynthesis}

\end{document}